\documentclass[prd,aps,twocolumn,preprintnumbers,amsmath,amssymb,nofootinbib,superscriptaddress,notitlepage]{revtex4}
\usepackage{graphicx}
\usepackage{epsfig}
\usepackage[utf8]{inputenc}
\usepackage{color}
\usepackage{epstopdf}
\usepackage[colorlinks=true,
linkcolor=blue,
breaklinks=true,
urlcolor=blue,
citecolor=blue]{hyperref}

\newcommand{\nn}{\nonumber}
\newcommand{\be}{\begin{equation}}
\newcommand{\ee}{\end{equation}}
\newcommand{\bea}{\begin{eqnarray}}
\newcommand{\eea}{\end{eqnarray}}
\newcommand{\ba}{\begin{array}}
\newcommand{\ea}{\end{array}}
\newcommand{\bi}{\begin{itemize}}
\newcommand{\ei}{\end{itemize}}

\renewcommand{\vec}[1]{\mbox{\boldmath $#1 \!\!$ \unboldmath}}

\newcommand{\lf}{\left}
\newcommand{\rg}{\right}

\newcommand{\lam}{\Lambda}

\newcommand{\dd}{\text{d}}
\newcommand{\eeBB}{e^+e^-\to B\overline B}

\newcommand{\CP}{\textit{CP }}

\newcommand{\ihep}{\affiliation{Institute of High Energy Physics, Chinese Academy of Sciences, Beijing 100049, China}}

\newcommand{\ucas}{\affiliation{University of Chinese Academy of Sciences, Beijing 100049, China}}

\newcommand{\imp}{\affiliation{Institute of Modern Physics, Chinese Academy of Sciences, Lanzhou 730000, China}}

\newcommand{\csr}{\affiliation{Research Center for Hadron and CSR Physics, Lanzhou University and Institute of Modern Physics of CAS, Lanzhou 730000, China}}

\begin{document}

\title{Production and decay of hyperons in a transversely polarized electron-positron collider}

\author{Xu Cao}\email{caoxu@impcas.ac.cn}
\imp
\ucas
\csr

\author{Yu-Tie Liang}\email{liangyt@impcas.ac.cn}
\imp
\ucas

\author{Rong-Gang Ping}\email{pingrg@mail.ihep.ac.cn}
\ihep
\ucas
\date{\today}

\begin{abstract}
  \rule{0ex}{3ex}
The self-polarization of relativistic electrons or positrons moving in a magnetic field at a storage ring occurs through the emission of spin-flip synchrotron radiation, known as the Sokolov-Ternov effect. The resulting transverse polarizations of the colliding electrons and positrons, away from the depolarization resonances, allow for precise investigation of the spin entangled hyperon-antihyperon pairs via virtual photon or charmonium decay. The feasibility study reveals a promising increase in the statistical sensitivity of the \CP violation signal after considering the transverse polarizations of the lepton beams.
\end{abstract}

\maketitle

\section{Introduction}
Transversely-polarized beams at an electron-positron collider are particularly interesting in the search for new sources of \CP violation through the measurement of \CP odd azimuthal asymmetry \cite{Ananthanarayan:2003wi}. They also offer a potential opportunity for revealing fundamental interactions and probing new physics \cite{Moortgat-Pick:2005jsx}. For instance, some of specific modulations of azimuthal distributions would originate from the interference of operators of new physics beyond the Standard Model (SM) with those in the SM \cite{Wen:2023xxc}. Additionally, transverse-polarization monitoring is a key technique for beam energy calibration using the resonant depolarization method, thereby improving the measurement of the mass and width of narrow resonances, such as the $Z-$ boson and $J/\psi$ \cite{Chen:2022rgo}.

The well-known Sokolov-Ternov effect \cite{Sokolov:1963zn} refers to the spin-flip processes through synchrotron radiation emission, which
results in a natural build-up of transverse polarization through the competition between radiation self-polarization and spin diffusion \cite{Baier:1965po,Jackson:1975qi}. In the mid-1970s, the ACO storage ring at Orsay \cite{Montague:1983yi} and the improved VEPP-2M at the Budker Institute of Nuclear Physics (BINP) \cite{Kurdadze:1975zm} both approached the theoretical limit of polarization levels, approximately $P_0 = {8}/{5\sqrt{3}} \simeq$ 92.4\%. At a beam energy of 3.7 GeV, SPEAR II found the equilibrium value of polarization to be about 0.76 \cite{Learned:1975sw}. The degree of transverse beam polarization at the CERN Large Electron Positron storage ring (LEP) was observed to be around 9.1\% \cite{Knudsen:1991cu}. In the future, it is possible to polarize the beams up to 10\% in a few hours at a designed energy of 45.6 GeV at the high-energy Circular Electron-Positron Collider (CEPC) \cite{Nikitin:2019nxl}. The same level of transverse polarization of the beams is expected in the Future Circular Collider (FCC-ee) \cite{Blondel:2019jmp}. As the first and only ring to provide longitudinal lepton spin polarization at high energy (27.5 GeV), the production of polarized electron and positron beams in HERA relied on the Sokolov-Ternov effect, where the spin rotators on either side of the interaction points converted the polarization of the beam from transverse to longitudinal, or vise versa. High luminosity at SuperKEKB prevents the use of the Sokolov-Ternov effect from accumulating  or yielding longitudinal polarization \cite{USBelleIIGroup:2022qro}.

The characteristic rise time for the Sokolov-Ternov polarization to build up from an unpolarized state is given by:
\be
\tau_{0} = \frac{8}{5\sqrt{3}} \frac{m_e^2 c^2 \rho^2 R}{e^2 \hbar \gamma^5}\,,
\ee
where the Lorentz factor is $\gamma = E_e/m_e$, leading to $\tau_{0} = 2.8$ hours at $E_e = 2.0$ GeV, using an average radius $R = 38$ meters and an effective radius $\rho = 9.3$ meters for BEPCII (Beijing Electron-Positron Collider).
The degree of transverse polarization $P_T =P_0 (1 - e^{-t/\tau_0})$ reaches 0.28 after one hour of beam injection, and it increases with the rise of beam energy over the same duration of injection time at the same accelerator.
The depolarization resonances from imperfections in the magnetic field occur at certain energies determined by gyromagnetic anomaly, leading into vanishing polarization of beams.
The realistic situation is, of course, more complicated due to many other sources of depolarization resonances. However, the degree of transverse polarization of the beams at BEPCII and future super $\tau$-charm facility (STCF) \cite{Achasov:2023gey}, away from the depolarization resonances, is expected to be of fair magnitude, resulting in a sizable impact on the angular distribution of final particles.

For a long time, spin asymmetries and correlations in hyperon-pair production in unpolarized electron-positron collisions have been proposed to measure simultaneously the electric-magnetic form factors and decay parameters of hyperons and anti-hyperons \cite{Czyz:2007wi,Chen:2007zzf,Faldt:2017kgy}. This approach allows for the investigation of \CP nonconservation parameters with high precision \cite{Lee:1957qs,Donoghue:1985ww,Donoghue:1986hh}.
Recently, \CP violation in hyperon decays at super-charm-tau factories with a longitudinally polarized electron beam has been investigated \cite{Salone:2022lpt,Zeng:2023wqw}, building upon earlier efforts \cite{Dubnickova:1992ii,Brodsky:2003gs,Tomasi-Gustafsson:2005svz}. However, the effects from transverse beam polarization were never taken into account in the analysis and simulation of the $e^+e^-\to B\overline B$ reactions, except for some earlier attempts on angular distributions \cite{Tsai:1975bd,Bletzacker:1976npa,Hikasa:1985qi}.
In particular, the hyperon transverse polarization in $\psi(3686) \rightarrow \Lambda\bar{\Lambda}$ \cite{BESIII:2023euh} and $ \Sigma^{+} \bar{\Sigma}^-$ decays \cite{BESIII:2020fqg} is expected to be affected by the transverse polarization of the lepton beams, as is the case for $\psi(3770)$ \cite{BESIII:2021cvv}. The transverse polarization of double-strange baryons $\Xi$ observed with unprecedented accuracy by the BESIII collaboration in $\psi(3686) \rightarrow \Xi^-\bar{\Xi}^+$ \cite{BESIII:2022lsz,Liu:2023xhg}, $\Xi^0\bar{\Xi}^0$ \cite{BESIII:2023lkg}, $\Xi(1530)^{-}\bar\Xi(1530)^{+}$ and $\Xi(1530)^{-}\bar\Xi^{+}$ \cite{BESIII:2019dve} also requires further scrutiny.
The polarization of most strange baryons $\Omega$ in $\psi(3686)\to \Omega^- \bar \Omega^+$ \cite{BESIII:2020lkm} deserves more attention if enough events are accumulated. Additionally, data from different isospin channels would be helpful for understanding the hyperon electromagnetic form factors \cite{Dai:2023vsw}.

In Section \ref{sec:anlytical}, we provide an analytical illustration of the effect of the transverse polarizations of the lepton beams on the production and decay of hyperons. The numerical outcome of moments analysis and statistical significance test is presented in Section \ref{sec:numerical}. Finally, we summarize the results to conclude the paper in Section \ref{sec:close}.

\section{Production and decay chains of hyperons} \label{sec:anlytical}
\subsubsection{Spin Density Matrix} \label{subsec:SDM}

As a start, we consider the annihilation of a particle-antiparticle pair $(f\bar f)$ to a virtual photon ($\gamma^*$) of energy squared $s$: $f(\lambda_1)\bar f(\lambda_2)\to \gamma^*(\lambda)$. In this process, annihilation conserves helicity, yielding $\lambda=\lambda_1-\lambda_2$. If the particle and antiparticle, both with mass $M$, are of the same spin 1/2, there are three helicity configurations: $(\lambda_1,\lambda_2)=(\pm 1/2,\pm 1/2), (-1/2,1/2)$ and $(1/2,-1/2)$. There are four helicity amplitudes $A_{\lambda_1,\lambda_2}$, but only two are independent, for example, $A_{1/2,1/2}=A_{-1/2,-1/2}= 2 \sqrt{2} M G_E$ and $A_{1/2,-1/2}=A_{-1/2,1/2}=2 \sqrt{ s} G_M$. Here, $G_{E,M}$ are the electromagnetic form factors of the particle.
If the particle (antiparticle) is the structureless electron (positron), the helicity of the electron and positron must be opposite; otherwise, the helicity amplitude with a vanishing helicity difference $\lambda_z=0$ is suppressed by a factor of $m_e/\sqrt{ s}$. As a result, the photon only couples right-handed particles to left-handed antiparticles and vice versa.

Due to synchrotron radiation when positrons and electrons circulate in the storage ring, the transition probabilities of the two spin projections of positrons and electrons, guided by the magnetic field in the storage ring, are different. This causes the spin orientation of positrons to tend toward the direction of the guiding magnetic field, while the orientation of electrons is opposite. Consequently, as the lepton beam remains in the storage ring for an extended period, they will acquire a transverse polarization $\mathcal{P}_t=p_x+i p_y=P_T e^{i\phi_+ }$ and $\bar{\mathcal{P}}_t=\bar{p}_x+i \bar{p}_y=\bar{P}_T e^{i\phi_- }$ for positron and electron, respectively, where $P_T=|\mathcal{P}_t|$. Here, $p_x (\bar{p}_x)$ represents the degree of transverse polarization in the scattering plane, and $p_y (\bar{p}_y)$ represents the degree of polarization perpendicular to the scattering plane \cite{Moortgat-Pick:2005jsx}.
The angles $\phi_+$ and $\phi_-$ represent the azimuthal angles of the positron and electron polarizations, respectively, with respect to the lab system. In a positron-electron annihilation experiment with symmetric beam energy, the Sokolov–Ternov effect requires the equal degree of polarization $P_T = \bar{P}_T$ and $\phi_+ = \phi_-= \pi/2$. This means that the positron and the electron have the same polarization vector in the individual helicity frame \footnote{In an $e^+e^-$ storage ring, the $z$-axes of the helicity frames for the electron and positron are aligned with their respective directions of motion, and they share the same $x$-axis along the radical direction, while the vertical $y$-axis is oriented in the opposite direction.}. The spin density matrix of the leptons is represented in their helicity frame as:
\bea \nn
\rho^- &=& \frac{1}{2}
\left(
\begin{array}{cc}
1 + \mathcal{P}_z & \mathcal{P}_t  \\[0.5em]
\mathcal{P}_t^* & 1 - \mathcal{P}_z \\
\end{array}
\right) \text{for } e^-\,,\\
\rho^+ &=& \frac{1}{2}
\left(
\begin{array}{cc}
1 + \bar{\mathcal{P}}_z& \mathcal{P}_t  \\[0.5em]
\mathcal{P}_t^* & 1 - \bar{\mathcal{P}}_z\\
\end{array}
\right) \text{for }e^+ \,,\nonumber
\label{eq:leptonhelicity}
\eea
in the most general case of considering both a longitudinal and a transverse component of polarization vectors.

In the laboratory system, for the process of $e^+ e^-\to \gamma^*/\psi$ annihilation, the spin density matrix (SDM) element of $\gamma^*/\psi$ is given by:
\bea
\rho_{m,m'}^{\gamma^*/\psi}&=&\sum_{\lambda_1,\lambda_1' \lambda_2,\lambda_2'}D_{m,\lambda_1-\lambda_2}^{1*} (0,0,0)D_{m',\lambda'_1-\lambda'_2}^1 (0,0,0)\nn\\
&\times&\rho_{\lambda_1,\lambda_1'}^+ \rho_{\lambda_2,\lambda_2'}^- \delta_{\lambda_1,-\lambda_2 } \delta_{\lambda_1',-\lambda_2'} \,,
\eea
where $\lambda_1, \lambda_1' (\lambda_2, \lambda_2')$ represent the helicity values of positron (electron). The Dirac $\delta-$function in the above equation ensures the conservation of helicity during the positron-electron annihilation process as discussed from the beginning of this subsection.
Performing a simple algebraic calculation on the above equation, we obtain:
\bea
\rho^{\gamma^*/\psi} &=& \frac{1}{2}
\left(
\begin{array}{ccc}
(1 - \mathcal{P}_z)(1 + \bar{\mathcal{P}}_z) & 0& P_T^2  \\[0.5em]
0&0&0\\
P_T^2 &0& (1 - \mathcal{P}_z)(1 + \bar{\mathcal{P}}_z) \\
\end{array}
\right), \nn
\eea

Based on the spin density matrix, we conduct a simple analysis of the polarization of $\gamma^*/\psi$. For a particle with spin $s=1$, its overall degree of polarization is defined by $d={1\over \sqrt{2s}} [(2s+1)\text{Tr}(\rho^{\psi2})-1]^{1/2}=\sqrt{1+3\mathcal{P}_z^2+3P_T^4 }/2$ by taking $\bar{\mathcal{P}}_z = 0$ for simplicity, indicating that the presence of $P_T$ increases the overall degree of polarization of $\gamma^*/\psi$ state as well as the $\mathcal{P}_z$. This polarization has two sources: one is the linear polarization $\mathcal{Q}=(q_x,q_y,q_z)$, and the other is the tensor polarization $T_{ij} (i,j=x,y,z)$. The spin density matrix elements of $\rho^\psi$ can be expressed using $\mathcal{Q}$ and $T_{ij}$
\begin{widetext}
\bea
\rho^{\gamma^*/\psi} &=& \frac{1}{3}
\left(
\begin{array}{ccc}
1+{3q_z\over2}+\sqrt{3\over2} (T_{xx}+T_{yy}+2T_{zz} ) & {3(q_x-iq_y)\over 2\sqrt{2}}& \sqrt{3\over2}(T_{xx}-T_{yy})  \\[0.5em]
{3(q_x+iq_y)\over 2\sqrt{2}}&1+\sqrt{6} T_{xx}+\sqrt{6} T_{yy}&{3(q_x-iq_y)\over 2\sqrt{2}}\\
 \sqrt{3\over2}(T_{xx}-T_{yy}) &{3(q_x+iq_y)\over 2\sqrt{2}}&  1-{3q_z\over2}+\sqrt{3\over2}(T_{xx}+T_{yy}+2T_{zz})\\
\end{array}
\right).
\eea
\end{widetext}
Comparing the elements of the $\rho^\psi$ matrix, we can obtain $q_x=q_y=0,~ q_z=-\mathcal{P}_z,~ T_{xx}={3P_t^2-1\over 2\sqrt{6}},~T_{yy}=-{1+3P_t^2 \over 2\sqrt{6}},~T_{zz}=1/\sqrt{6}$ with other $T_{ij} (i\neq j)=0$. It can be seen that the tensor polarization of $\rho^{\gamma^*/\psi}$ comes from the spin correlation and transverse polarization of the initial lepton beams, with linear polarization being solely from the longitudinal polarization of  beams.
In the following paper, we focus solely on the transverse polarization of the initial lepton beams. The study of a longitudinally polarized electron beam has been recently addressed independently in other studies \cite{Salone:2022lpt,Zeng:2023wqw}.

\subsubsection{\texorpdfstring{$e^+e^- \to \gamma^*/\psi \to \Lambda (p \pi^-) \bar\Lambda (\overline p \pi^+)$}{}} \label{sec:Lambda}

\begin{figure}[b]
    \centering
    \includegraphics[width=0.95\linewidth]{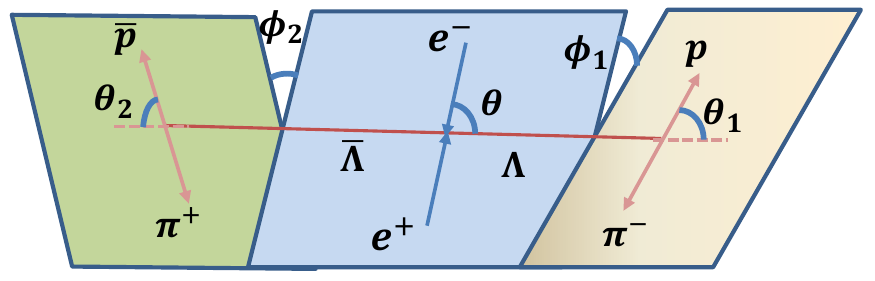}
    \caption{Helicity frame defined for the $\psi \to \Lambda(p\pi^-)
\overline\Lambda(\bar p\pi^+)$.}
    \label{fig:frame}
\end{figure}

In the laboratory system, the decayed $\Lambda$ particle moves in the direction defined by polar and azimuthal angles $(\theta,\phi)$. We calculate the joint angular distribution in the $\Lambda$ helicity system, as illustrated in Fig. \ref{fig:frame}. In this system, the $z$-axis aligns with the $\Lambda$ particle's direction of motion, the $y$-axis is perpendicular to the $\Lambda$ production plane, i.e., $\hat y=\hat p_+\times \hat{p}_\Lambda$, where $\hat p_+$ and $\hat{p}_\Lambda$ are the unit momentum vectors for the positron and $\Lambda$, respectively. The $x$-axis lies in the $\Lambda$ production plane, forming a right-handed $x$-$y$-$z$ coordinate system. After boosting the momenta of $p$ and $\pi^-$ in laboratory frame to the $\Lambda$ rest frame, they define the $\Lambda$ decay plane together with the momenta of $\Lambda$ in laboratory frame. The angle between the proton momentum in the $\Lambda$ decay plane and $\Lambda$ momentum in laboratory frame is defined as $\theta_1$, and the angle between the $\Lambda$ production and decay plane is defined as $\phi_1$. Similar helicity angles $(\theta_2,\phi_2)$ are defined in the same manner.
In the $\Lambda$ helicity system, the $J/\psi$ SDM is calculated by transforming it from the laboratory system to this helicity system, as follows:
\bea
  \rho_1^{i,j}(\theta,\phi) &\equiv& \sum\limits_{k,k^\prime=\pm 1}  \rho^{\gamma^*/\psi}_{k,k^\prime} {\cal D}_{k,i}^{1*}(\phi,\theta,0) \, {\cal D}^1_{k^\prime,j}(\phi,\theta,0)
  \nn\\ &=& \sum\limits_{k=\pm 1} \lf[
  {\cal D}_{k,i}^{1*}(\phi,\theta,0) \, {\cal D}^1_{k,j}(\phi,\theta,0) \rg. \qquad \qquad \qquad \nn \\
   &+& \lf. P_T^2 {\cal D}_{k,i}^{1*}(\phi,\theta,0) \, {\cal D}^1_{-k,j}(\phi,\theta,0) \rg]\, .
  \label{eq:defrho0tran}
\eea
Thus the effects from transverse beam polarization occur only if both beams are polarized and generate interference terms between left- and right-helicity amplitudes \cite{Hikasa:1985qi}.
The explicit form of the reduced $\rho_1$ is given by
\begin{widetext}
\bea
\rho_1(\theta,\phi) &=& \frac{1}{2} \cdot
\left(
\begin{array}{ccc}
 \frac{1+\cos^2\!\theta}{2} &
-\frac{\cos\theta\sin\theta}{\sqrt{2}} & \frac{\sin ^2\!\theta}{2} \\[0.5em]
-\frac{\cos\theta\sin\theta}{\sqrt{2}} &
\sin ^2\!\theta & \frac{\cos\theta\sin\theta}{\sqrt{2}} \\[0.5em]
\frac{\sin ^2\!\theta}{2} & \frac{\cos \theta \sin \theta}{\sqrt{2}}  &
\frac{1+\cos^2\!\theta}{2} \\
\end{array}
\right)
\nn \\ &+& \frac{1}{2} P_T^2 \cdot
\left(
\begin{array}{ccc}
\frac{\sin^2\theta}{2} \cos2 \phi&
\frac{\cos\theta\sin\theta}{\sqrt{2}} \cos2 \phi - i \frac{\sin\theta}{\sqrt{2}} \sin2 \phi&
\frac{1+\cos^2\!\theta}{ 2} \cos2 \phi - i \cos\theta \sin2 \phi \\[0.5em]
\frac{\cos\theta\sin\theta}{\sqrt{2}} \cos2 \phi + i \frac{\sin\theta}{\sqrt{2}} \sin2 \phi &
-\sin^2\theta \cos2 \phi &
-\frac{\cos\theta\sin\theta}{\sqrt{2}} \cos2 \phi + i \frac{\sin\theta}{\sqrt{2}} \sin2 \phi \\[0.5em]
\frac{1+\cos^2\!\theta}{2} \cos2 \phi + i \cos\theta \sin2 \phi &
-\frac{\cos\theta\sin\theta}{\sqrt{2}} \cos2 \phi - i \frac{\sin\theta}{\sqrt{2}} \sin2 \phi &
\frac{\sin^2\theta}{2} \cos2 \phi \\
\end{array}
\right)
\,. \qquad
\label{eq:SDMtrans}
\eea
\end{widetext}

The density matrix for the production process is the sum of the contributions from the two helicities~\cite{Perotti:2018wxm,Salone:2022lpt}:
\be
\rho^{\lambda_1,\lambda_2;\lambda_1',\lambda_2'}_{B\overline B} \propto A_{\lambda_1,\lambda_2} \, A^*_{\lambda'_1,\lambda'_2} \, \rho_1^{\lambda_1-\lambda_2,\lambda'_1-\lambda'_2} \,, 
  \label{eqn:amp}
\ee
with the helicity amplitudes of the photon transition to a pair of baryon-antibaryon $A_{1/2,1/2}=A_{-1/2,-1/2} = \sqrt{(1-\alpha_\psi)/2}$ and  $A_{1/2,-1/2}=A_{-1/2,1/2} = \sqrt{1+\alpha_\psi} \, e^{-i\Delta\Phi}$.
Here $\alpha_\psi = (M_\psi^2 |G^\psi_{M}|^2 - 4 M_B^2|G^\psi_{E}|^2)/(M_\psi^2 |G^\psi_{M}|^2 + 4 M_B^2|G^\psi_{E}|^2)$ is the decay parameter of charmonium to baryon-antibaryon, and the $\Delta\Phi = \arg(G^\psi_{E}/G^\psi_{M})$ is the relative phase between $\psi$ electric and magnetic form factors.
The general expression for the joint density matrix of the $B\overline B$ pair is:
\begin{equation}
\rho_{B\overline B}=\sum_{\mu,\nu=0}^{3}C_{\mu\nu}\, \sigma_\mu^{B}\otimes{\sigma}_{\nu}^{\overline B}\ ,
\label{eqn:sig12}
\end{equation}
where a set of four Pauli matrices $\sigma_\mu^{B}(\sigma_\nu^{\overline B})$ in the  $B({\overline B})$ rest frame is used and $C_{\mu\nu} = C_{\mu\nu}^U + C_{\mu\nu}^T$ is a $4\times 4$ real matrix representing polarizations and spin correlations of the baryons.
The elements of  the $C_{\mu\nu}$ matrix are functions of the production angle $\Omega(\theta,\phi)$ of the $B$ baryon:
\begin{widetext}
\bea
(C_{\mu\nu})&=& \frac{3}{3+\alpha_\psi}
\cdot\left(
\begin{array}{cccc}
 1\!+\!\alpha_\psi  \cos^2\!\theta & 0 & {\beta_\psi  {\sin\!\theta \cos\!\theta}} & 0 \\
 0 & \sin ^2\!\theta & 0 & {\gamma_\psi  {\sin\!\theta \cos\!\theta}}  \\
 -{\beta_\psi  {\sin\!\theta \cos\!\theta}}  & 0 & \alpha_\psi  \sin^2\!\theta  & 0 \\
 0 & -{\gamma_\psi  {\sin\!\theta \cos\!\theta}}  &  0 & -\alpha_\psi\!-\!\cos ^2\!\theta \\
\end{array}
\right) \nn \\
&+& \frac{3 P_T^2}{3+\alpha_\psi}
\cdot\left(
\begin{array}{cccc}
 \alpha_\psi \sin^2\!\theta \cos\!2 \phi & -{\beta_\psi  {\sin\!\theta}} \sin\!2\phi & -{\beta_\psi {\sin\!\theta \cos\!\theta}} \cos\!2\phi & 0 \\
 -{\beta_\psi {\sin\!\theta}} \sin\!2\phi & (\alpha_\psi + \cos^2\!\theta) \cos\!2\phi & -(1 + \alpha_\psi) \cos\!\theta \sin\!2\phi &  -{\gamma_\psi {\sin\!\theta \cos\!\theta}} \cos\!2\phi \\
 {\beta_\psi {\sin\!\theta \cos\!\theta}} \cos\!2\phi  & (1 + \alpha_\psi) \cos\!\theta \sin\!2 \phi & (1 + \alpha_\psi \cos^2\!\theta) \cos\!2 \phi  & -{\gamma_\psi {\sin\!\theta}} \sin\!2\phi \\
 0 & {\gamma_\psi {\sin\!\theta \cos\!\theta}} \cos\!2\phi &  -{\gamma_\psi {\sin\!\theta \sin\!2\phi}}  & - \sin^2\!\theta \cos\!2\phi \\
\end{array}
\right),\label{eq:cmatrixtran}
\eea
\end{widetext}
with $\beta_\psi = \sqrt{1-\alpha_\psi^2} \sin\!\Delta\Phi$ and $\gamma_\psi = \sqrt{1-\alpha_\psi^2} \cos\!\Delta\Phi$.

Therefore a transverse polarization of the final state baryon is only allowed in the direction normal to the plane spanned by the incoming beam and the outgoing baryon:
\bea
P_y^B &=& \frac{\beta_\psi  {\sin\!\theta \cos\!\theta} (1 - P_T^2 \cos\!2 \phi)}{1+\alpha_\psi\cos^2\!\theta + \alpha_\psi P_T^2 \sin^2\!\theta \cos\!2 \phi} \,, \\
P_x^B &=& \frac{-P_T^2 \beta_\psi {\sin\!\theta} \sin\!2 \phi}{1+\alpha_\psi\cos^2\!\theta + \alpha_\psi P_T^2 \sin^2\!\theta \cos\!2 \phi} \,,
\eea
with vanishing longitudinal polarization component. Besides, the
transverse polarization of beams introduces two new correlations $C_{xy} = -C_{yx}$, $C_{yz} = C_{zy}$ in addition to the four existing ones, which further constrain the decay parameters of hyperons, and thus provides more stringent tests of the \CP violation.

\begin{figure}[t]
    \centering
    \includegraphics[width=\linewidth]{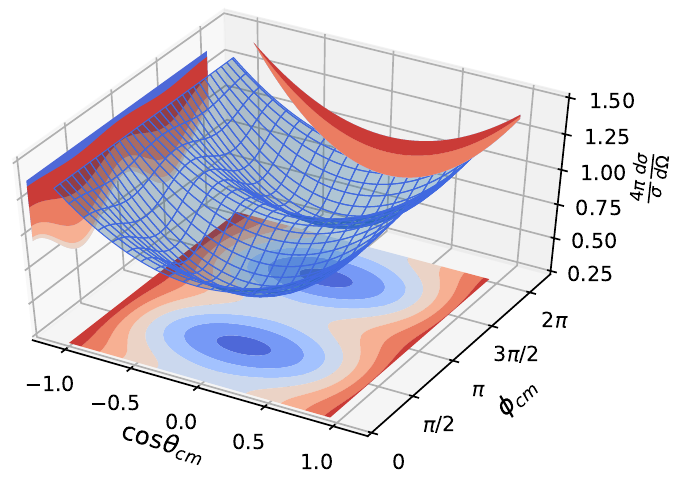}
    \caption{The three dimenional plot and the projection of angular distribution of $\Lambda$ hyperon in $e^+ e^- \to \psi(3686) \to \Lambda \overline\Lambda$ with $\alpha = 0.69$ and $\Delta\Phi = 23^\circ$ \cite{BESIII:2023euh}, and $P_T = 0.5$.
    }
    \label{fig:dcsPysimul}
\end{figure}

The baryon angular distribution is
\be
    \frac{4\pi}{\sigma}\frac{\dd\sigma}{\dd\Omega_B}=
    \frac{3}{3+\alpha_\psi} ( 1+\alpha_\psi\cos^2\!\theta + \alpha_\psi P_T^2 \sin^2\!\theta \cos2 \phi ) \,
    \label{eq:dSigdOm}
\ee
and Fig. \ref{fig:dcsPysimul} shows its three dimensional distribution.
So that if the transverse polarization of the final-state particles is not measured, the effects of transverse polarizations are absent in the $\phi$-averaged cross section, albeit, present in the $\theta$-averaged cross section:
\be
    \frac{2\pi}{\sigma}\frac{\dd\sigma}{\dd\phi}=
    -2\cos\theta_0\frac{ 3+\alpha_\psi\cos^2\!\theta_0 + \alpha_\psi P_T^2 \cos2 \phi (3-\cos^2\!\theta_0)}{3+\alpha_\psi} \,, \label{eq:dSigdphi}
\ee
with $(\theta_0, \pi-\theta_0)$ being the
detector coverage of the solid angle around the interaction point (IP).
Therefore the degree of transverse polarization would be measured through $e^+ e^- \to \mu^+ \mu^-$ and $e^+ e^-$ azimuth angular distributions.
The distribution of scattering angle has been previously explored in BES \cite{BES:1995wyo}, BESIII \cite{BESIII:2018wid} and KEDR at the VEPP-4M \cite{Anashin:2018iwp}, but the azimuth distributions did not receive any attention.

There are five global parameters to describe a process $\eeBB$ followed by single-step weak two-body decays of the hyperon $B$ and the antihyperon $\overline B$~\cite{Perotti:2018wxm}.
For decay $B\!\to\! b\pi$ and the corresponding charge conjugate (c.c.) decay mode $\overline B\!\to\! \overline b\overline\pi$, like $e^+e^-\to \gamma^*/\psi \to\Lambda(p \pi^-) \overline\Lambda(\overline p \pi^+)$,
they are represented by the vector $\boldsymbol{\omega} \equiv (\alpha_\psi,\Delta\Phi,\alpha_-, {\alpha_+})$ with \textit{a priori} known $P_T^2$ and $\alpha_-$ (or ${\alpha_+}$) being the decay parameter of $B\!\to\! b\pi$ (or $\overline B\!\to\! \overline b\overline\pi$).
The joint angular distribution $\mathcal{W}(\boldsymbol{\xi})$ can be expressed with respect to the vector $\boldsymbol{\xi}\equiv(\Omega_B,\Omega_{b},\Omega_{\overline{b}})$ representing a complete set of the kinematic variables describing a single-event configuration in the six dimensional phase space \cite{Hong:2023soc,Zeng:2023wqw}:
\begin{align}
\begin{gathered}
 \mathcal{W}(\boldsymbol{\xi})= \mathcal{F}_{0}+\beta_{\psi} ({\alpha_+} \mathcal{F}_{3}-\alpha_- \mathcal{F}_{4})\\\hfill
 +\alpha_-{\alpha_+}(\mathcal{F}_{1}+ \gamma_{\psi} \mathcal{F}_{2}+\alpha_{\psi} \mathcal{F}_{5}) \,,
 \label{eq:joint2PT}
\end{gathered}
\end{align}
where the angular function $\mathcal{F}_{i}(\xi)$~($i$ = 0, 1, ..., 5) are defined as
\begin{widetext}
\be
\begin{aligned}
\mathcal{F}_{0}=&1+\alpha_{\psi}\cos^{2}\theta + \alpha_\psi P_T^2 \sin^2\!\theta \cos2 \phi \,,\\
\mathcal{F}_{1}=&(\sin^{2}\theta + P_T^2 \cos2 \phi \cos^{2}\theta) \sin\theta_{1}\cos\phi_{1}\sin\theta_{2}\cos\phi_{2}
\\ & -(\cos^2\!\theta + P_T^2 \cos2 \phi \sin^2\!\theta ) \cos\theta_{1}\cos\theta_{2} \\ &
+ P_T^2 \sin\theta_{1} \sin\theta_{2} (\sin2 \phi \cos\theta \sin(\phi_{1}-\phi_{2}) + \cos2 \phi \sin\phi_{1}\sin\phi_{2}) \,,
\\ \mathcal{F}_{2}=&(1 - P_T^2 \cos2 \phi) \sin\theta \cos\theta(\sin\theta_{1}\cos\theta_{2}\cos\phi_{1}-\cos\theta_{1}\sin\theta_{2}\cos\phi_{2}) \\ & -
P_T^2 \sin2 \phi \sin\theta (\sin\theta_{1}\cos\theta_{2}\sin\phi_{1} + \cos\theta_{1}\sin\theta_{2}\sin\phi_{2}) \,, \\
\mathcal{F}_{3}=&(1 - P_T^2 \cos2 \phi) \sin\theta \cos\theta \sin\theta_{2}\sin\phi_{2} - P_T^2 \sin2 \phi \sin\theta {\sin\theta_2} \cos\phi_{2} \,, \\
\mathcal{F}_{4}=&(1 - P_T^2 \cos2 \phi) \sin\theta \cos\theta \sin\theta_{1}\sin\phi_{1} + P_T^2 \sin2 \phi \sin\theta {\sin\theta_1} \cos\phi_{1} \,, \\
\mathcal{F}_{5}=& (\sin^{2}\theta + P_T^2 \cos2 \phi \cos^{2}\theta )\sin\theta_{1}\sin\phi_{1}\sin\theta_{2}\sin\phi_{2} -\cos\theta_{1}\cos\theta_{2} \\ &
+ P_T^2 \sin\theta_{1} \sin\theta_{2}[ \sin2 \phi \cos\theta \sin(\phi_{1}-\phi_{2}) + \cos2 \phi \cos\phi_{1}\cos\phi_{2}] \,,
 \label{eq:joint2ang}
\end{aligned}
\ee
\end{widetext}
where the $\Omega_{b} (\theta_{1}, \phi_{1}$) (or $\Omega_{\overline{b}} (\theta_{2}, \phi_{2}$) are the
spherical coordinates of $b$ (or $\overline b$) relative to $B$ (or $\overline B$) in the helicity frame of $B$ (or $\overline B$).
The {\it helicity angles} are used here to parameterize the multidimensional phase space, which are in following the angular convention with those previous works \cite{Salone:2022lpt,Zeng:2023wqw}.

\begin{figure}[t]
    \centering
    \includegraphics[width=\linewidth]{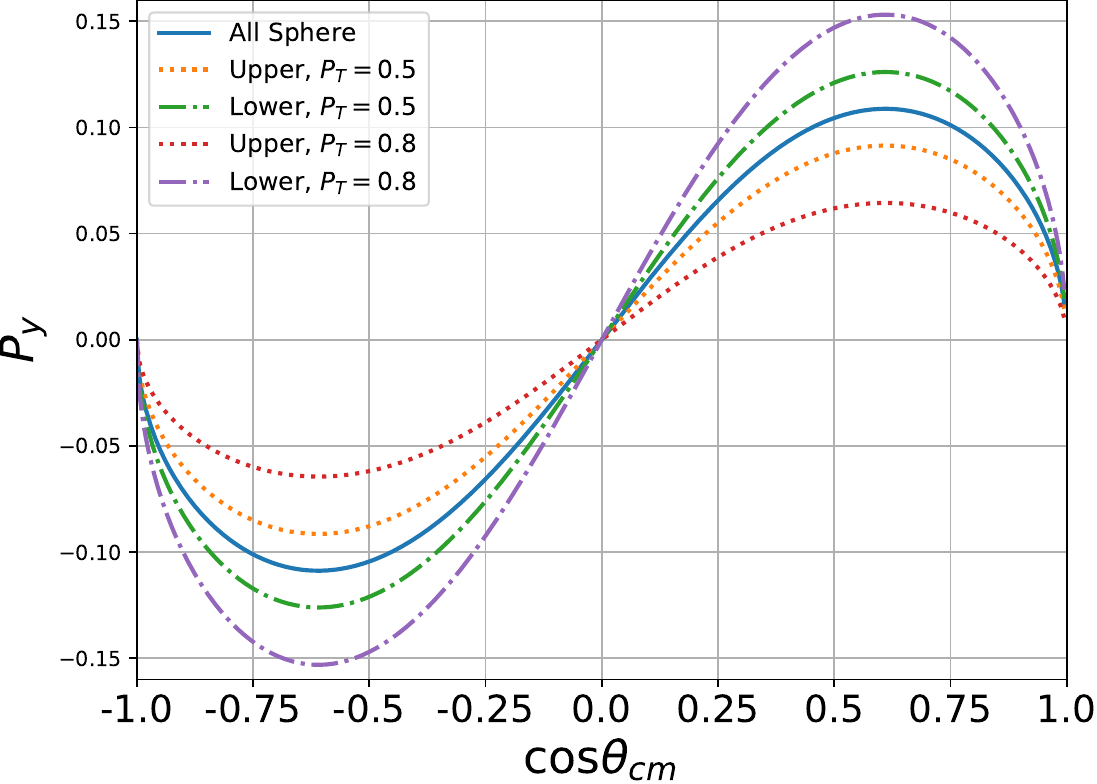}
    \includegraphics[width=\linewidth]{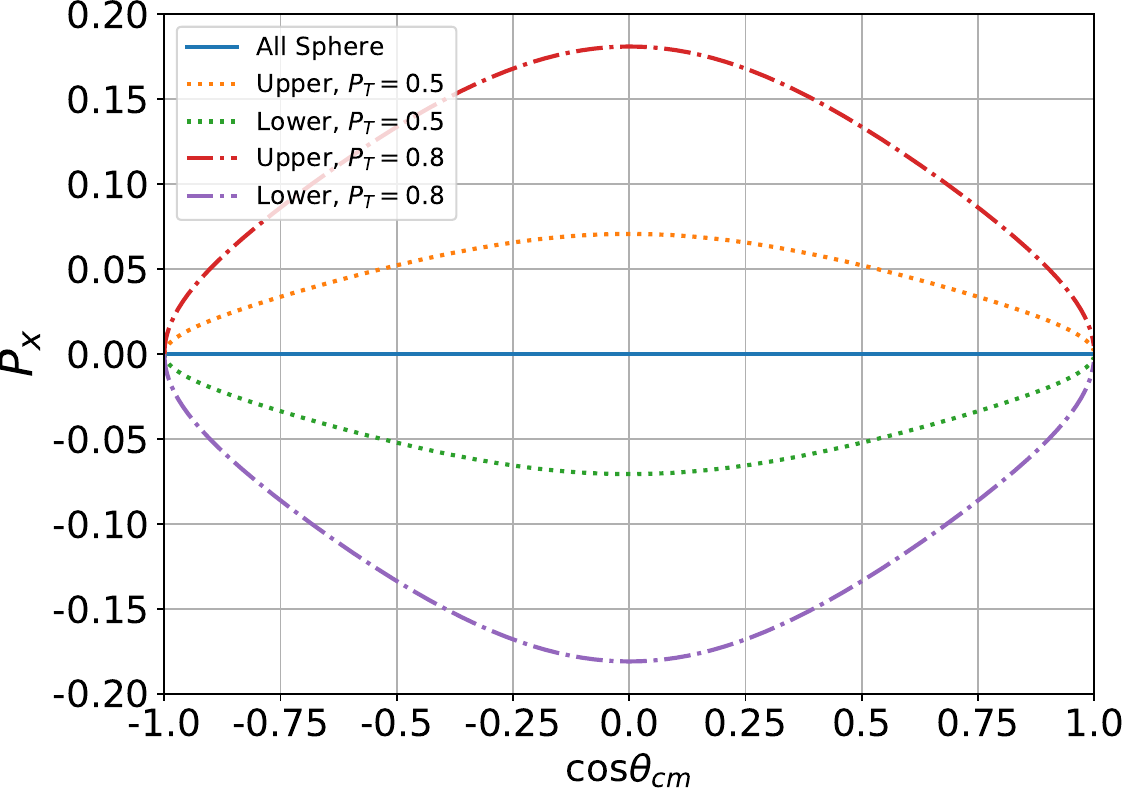}
    \caption{The transverse polarization $P_y$ and $P_x$ of $\Lambda$ hyperon in all azimuthal angles in comparison of upper and lower spheres. The parameters of $e^+ e^- \to \psi(3686) \to \Lambda \overline\Lambda$ with $\alpha_\psi = 0.69$ and $\Delta\Phi = 23^\circ$ \cite{BESIII:2023euh} are used.
    }
    \label{fig:semisphere}
\end{figure}

One can always define corresponding asymmetries in terms of events to select the $\cos2 \phi$ modulation as
\bea
\mathcal{W}^+_{\cos2 \phi}(\xi) &=& \int_{0}^{\pi/4} + \int_{3\pi/4}^{5\pi/4} + \int_{7\pi/4}^{2\pi} \mathcal{W}(\xi) \dd\phi \,, \\
\mathcal{W}^-_{\cos2 \phi}(\xi) &=&  \int_{\pi/4}^{3\pi/4} + \int_{5\pi/4}^{7\pi/4} \mathcal{W}(\xi) \dd\phi \,,
\eea
the latter of which is corresponding to the number of events in the range of upper semi-sphere $45^\circ$ to $135^\circ$ and $225^\circ$ to $315^\circ$, and the former is the number of events in the remaining lower semi-sphere.
Then $\mathcal{W}^+(\xi) - \mathcal{W}^-(\xi) $ is proportional to those terms of $P_T^2 \cos2 \phi$ dependence and $\mathcal{W}^+(\xi) + \mathcal{W}^-(\xi) $ integrates out the $P_T^2$ terms which are of $\cos2 \phi$ modulation.
To select the $\sin2 \phi$ modulation:
\bea
\mathcal{W}^+_{\sin2 \phi}(\xi) &=& \int_{0}^{\pi/2} + \int_{\pi}^{3\pi/2} \mathcal{W}(\xi) \dd\phi \,, \\
\mathcal{W}^-_{\sin2 \phi}(\xi) &=&  \int_{\pi}^{\pi/2} + \int_{3\pi/2}^{2\pi} \mathcal{W}(\xi) \dd\phi \,,
\eea
the former of which is corresponding to the number of events in the range of upper semi-sphere $0^\circ$ to $90^\circ$ and $180^\circ$ to $270^\circ$, and the latter is the number of events in the remaining lower semi-sphere.
Fig.~\ref{fig:semisphere} shows the transverse polarization $P_y$ and $P_x$ of baryon in all azimuthal angles in comparison of upper and lower spheres.
Instead, one can investigate the moments of the joint angular distributions as shown in Sec. \ref{sec:numerical}.

If identifying the decay chain of the hyperon with summation over the antihyperon spin directions, so called single tag events:
\begin{align}
\begin{gathered}
 \mathcal{W}(\boldsymbol{\xi})= \mathcal{F}_{0}-\beta_{\psi} \alpha_- \mathcal{F}_{4} \,, \label{eq:joint2tag}
\end{gathered}
\end{align}
which is useful to increase the statistical events if the transverse polarization of beams is of small degree.

\subsubsection{\texorpdfstring{$e^+e^- \to \gamma^*/\psi \to \Xi^-(\Lambda \pi^-)\bar\Xi^+(\overline{\Lambda} \pi^+)$}{}} \label{sec:xi}

\begin{figure}[b]
    \centering
    \includegraphics[width=0.95\linewidth]{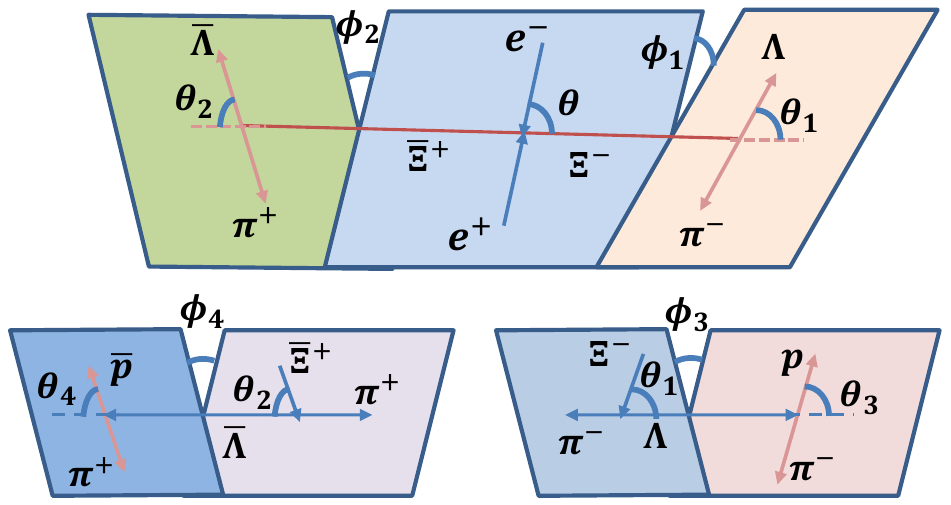}
    \caption{Upper panel: Helicity frame defined for the $\psi \to \Xi^-(\Lambda\pi^-)\bar\Xi^+(\bar\Lambda\pi^+)$.
    Lower panel: Helicity frame defined for $\Lambda(\bar\Lambda) \to p \pi^-(\bar p\pi^+)$.}
    \label{fig:frameXi}
\end{figure}

The definition of the helicity system for the first two decays in the processes $e^+e^- \to \gamma^*/\psi \to \Xi^- \bar\Xi^+$ with $\Xi \to\Lambda \pi$ and $\bar\Xi\to\bar\Lambda\pi^+$ is similar to that of $e^+e^-\to\psi\to\Lambda(p\pi^-)\bar\Lambda(\bar p\pi^+)$, as shown in Fig. \ref{fig:frameXi}. For the subsequent decays of $\Lambda(\bar\Lambda) \to p \pi^-(\bar p\pi^+)$, the polar angle $\theta_3(\theta_4)$ is defined as the angle between the momentum of the proton (anti-proton) and $\Lambda(\bar\Lambda)$ in the respective mother rest frame, and the azimuthal angle $\phi_3(\phi_4)$ is defined as the angle between the $\Lambda(\bar\Lambda)$ production and decay plane. The vector $\boldsymbol{\xi}:=(\theta,\phi,\theta_1,\phi_1,\theta_2,\phi_2,\theta_3,\phi_3,\theta_4,\phi_4)$ represents a complete set of the kinematic variables describing a single-event configuration in the ten-dimensional phase space, in line with the above definition of spherical coordinates in the helicity systems.

There are six global parameters to describe the complete angular distribution, represented by the vector $\boldsymbol{\omega} \equiv (\alpha_\psi,\Delta\Phi,\alpha_{1}, \overline{\alpha}_{2},\alpha_{3}, \overline{\alpha}_{4})$.
The joint angular distribution reads:
\begin{align}
\begin{gathered}
 \mathcal{W}(\boldsymbol{\xi}) =
 {\cal D}^0_\Xi \overline{\cal D}^0_\Xi C_{00} + {\cal D}^1_\Xi \overline{\cal D}^1_\Xi C_{{xx}} + {\cal D}^2_\Xi \overline{\cal D}^2_\Xi C_{{yy}} + {\cal D}^3_\Xi \overline{\cal D}^3_\Xi C_{{zz}} \\\hfill
 + ({\cal D}^1_\Xi \overline{\cal D}^2_\Xi - {\cal D}^2_\Xi \overline{\cal D}^1_\Xi ) C_{{xy}} + ({\cal D}^1_\Xi \overline{\cal D}^3_\Xi - {\cal D}^3_\Xi \overline{\cal D}^1_\Xi) C_{{xz}} \\\hfill + ({\cal D}^2_\Xi \overline{\cal D}^3_\Xi + {\cal D}^3_\Xi \overline{\cal D}^2_\Xi) C_{{yz}} + ({\cal D}^0_\Xi \overline{\cal D}^1_\Xi + {\cal D}^1_\Xi \overline{\cal D}^0_\Xi) P_x \\\hfill
+ ({\cal D}^0_\Xi \overline{\cal D}^2_\Xi - {\cal D}^2_\Xi \overline{\cal D}^0_\Xi) P_y \,,
 \label{eq:joint3}
\end{gathered}
\end{align}
with vanishing $P_z$ here. The parameters $C_{ij}$ and $P_i$($i=0$ or $x, y, z$) are given in Eq.~(\ref{eq:cmatrixtran}) with the substitution of $\alpha_\psi$ and $\Delta\Phi$ for those of $\gamma^*/\psi \to \Xi^- \bar\Xi^+$, and ${\cal D}^\mu_\Xi (\Omega_{\Xi} (\theta_{1}, \phi_{1}),\Omega_{\Lambda} (\theta_{3}, \phi_{3}))$ is the decay matrix of $\Xi \to\Lambda \pi$, $\Lambda \to p \pi^-$,
and $\overline{\cal D}^\nu_\Xi (\Omega_{\overline\Xi} (\theta_{2}, \phi_{2}),\Omega_{\overline\Lambda} (\theta_{4}, \phi_{4}))$ is for $\overline\Xi \to \overline\Lambda \pi$, $\overline\Lambda \to \overline p \pi^-$, respectively.
The ${\cal D}^\mu_\Xi$ matrix elements are explicitly written as \cite{Perotti:2018wxm}:
\begin{widetext}
\bea \nn
(D_\Xi^\mu)=\left(
\begin{array}{c}
 1\!+ \alpha_1 \alpha_3 \cos\!\theta_3 \\
 \alpha_1 \sin\!\theta_1 \cos\!\phi_1  + \alpha_3 [ \sin\!\theta_1 \cos\!\phi_1 \cos\!\theta_3 - \sin\!\phi_1 \sin\!\theta_3 (\beta_1 \cos\!\phi_3  + \gamma_1 \sin\!\phi_3) +
 \cos\!\theta_1 \cos\!\phi_1 \sin\!\theta_3 ( \gamma_1 \cos\!\phi_3 - \beta_1 \sin\!\phi_3) ] \\
 \alpha_1 \sin\!\theta_1 \sin\!\phi_1 + \alpha _3 [\sin\!\theta_1 \sin\!\phi_1 \cos\!\theta_3 + \cos\!\phi_1 \sin\!\theta_3 (\beta_1 \cos\!\phi_3  + \gamma_1 \sin\!\phi_3) +
 \cos\!\theta_1 \sin\!\phi_1 \sin\!\theta_3 ( \gamma_1 \cos\!\phi_3 - \beta_1 \sin\!\phi_3)]  \\
 \alpha_1\cos\!\theta_1 + \alpha_3 [\cos\!\theta _1 \cos\!\theta_3 -  \sin\!\theta_1 \sin\!\theta_3 (\gamma_1 \cos\!\phi_3 - \beta_1 \sin\!\phi_3)]
\end{array}
\right) \,, \qquad
\eea
with the substitution of $\{1,3\} \to \{2,4\}$ for $\overline{\cal D}^\nu_\Xi$.
This joint angular distribution, without explicit consideration of transverse polarization of beams, has been previously calculated in the literature. \cite{Hong:2023soc,Chen:2019hqi}.
For simplicity we demonstrate those joint angular distributions of single tag events with summation over the $\overline\Xi$ antihyperon spin directions:
\bea
 \mathcal{W}(\boldsymbol{\xi}) &=& (1+\alpha_{\psi}\cos^{2}\theta + \alpha_\psi P_T^2 \sin^2\!\theta \cos2 \phi ) (1 +\alpha_{\Xi}\alpha_{\lam} \cos\!\theta_{3}) \nn \\
 &-& P_T^2 \sin2 \phi \, \beta_{\psi}\sin\theta \{ \alpha_{\Xi} \sin\!\theta_1 \cos\!\phi_1 \nn \\ && + \alpha_{\lam} [ \sin\!\theta_1 \cos\!\phi_1 \cos\!\theta_3 - \sin\!\phi_1 \sin\!\theta_3 (\beta_{\Xi} \cos\!\phi_3  + \gamma_{\Xi} \sin\!\phi_3) +
 \cos\!\theta_1 \cos\!\phi_1 \sin\!\theta_3 ( \gamma_{\Xi} \cos\!\phi_3 - \beta_{\Xi} \sin\!\phi_3) ] \} \nn \\
 &-& (1 - P_T^2 \cos2 \phi) \beta_{\psi} \sin\theta \cos\theta \{ \alpha_{\Xi} \sin\!\theta_1 \sin\!\phi_1 \nn \\ && + \alpha_{\lam} [\sin\!\theta_1 \sin\!\phi_1 \cos\!\theta_3 + \cos\!\phi_1 \sin\!\theta_3 (\beta_{\Xi} \cos\!\phi_3  + \gamma_{\Xi} \sin\!\phi_3) +
 \cos\!\theta_1 \sin\!\phi_1 \sin\!\theta_3 ( \gamma_{\Xi} \cos\!\phi_3 - \beta_{\Xi} \sin\!\phi_3)] \} \,,
 \label{eq:joint3tag}
\eea
\end{widetext}
After integrating out the $P_T$ terms \cite{Faldt:2017yqt,Chen:2019hqi} it is used in the measurement of $\Lambda_c$ case \cite{BESIII:2019odb}.

\section{Discussions} \label{sec:numerical}

At the unpolarized electron-positron collider, the parasitic production of transverse polarization of beams provides new degrees of freedom for physical research.
Compared to the unpolarized beams, the formulas describing particle production and decay become more complex, but on the other hand, they provide us with more observational degrees of freedom to study the dynamics of decay processes.
The transverse polarization effect, in addition to being prominently expressed in the angular distribution of final-state particles, can also be manifested in the moment distribution of particles at various decay levels.
Furthermore, the presence of transverse polarization provides additional polarization information for measuring the asymmetry parameters of hyperon decay, which is beneficial for improving measurement accuracy, for instance, the \CP violation parameters as shown in the following analysis.

\subsection{Moments analysis} \label{sec:moments}

For instance, we construct the following observables by using the angles $\theta_1, \theta_2, \phi_1$, and $\phi_2$ detected in the process of $e^+ e^-\to\psi(3686)\to\Lambda\bar\Lambda,\Lambda\to p\pi^-,\bar\Lambda\to \bar p\pi^+$,
\begin{eqnarray}
\mu_1&=&\sin\theta_1\sin\theta_2[\sin(2\phi)\cos\theta\sin(\phi_1-\phi_2)\nonumber\\
     &+& \cos(2\phi)\sin\phi_1\sin\phi_2 ],\nonumber\\
\mu_2&=&\cos(2\phi)\sin\theta\cos\theta[\sin\theta_1\cos\theta_2\cos\phi_1\nonumber\\
     &-& \cos\theta_1\sin\theta_2\cos\phi_2],\\
\mu_3&=&\sin(2\phi)\sin\theta\cos\phi_2,\nonumber\\
\mu_4&=&\sin(2\phi)\sin\theta\cos\phi_1,\nonumber\\
\mu_5 &=&\sin\theta_1\sin\theta_2[\sin(2\phi)\cos\theta\sin(\phi_1-\phi_2)\nonumber\\
      &-& \cos(2\phi)\cos\phi_1\cos\phi_2]\nonumber.
\end{eqnarray}

The moments derived from these observables represent taking their averages over the joint angular distribution. Their moments with respect to the $\cos\theta$ angular distribution are expressed as:
\begin{equation}
{d\langle \mu_i \rangle\over d\cos\theta} ={\int \mathcal{W}({\bf\xi})\mu_i d\cos\theta_1d\cos\theta_2d\phi_1d\phi_2 \over  \int \mathcal{W}({\bf\xi}) d\cos\theta_1d\cos\theta_2d\phi_1d\phi_2}, (i=1,2,..,5).
\end{equation}
Then one has
\begin{eqnarray}\label{moments}
{d\langle \mu_1 \rangle\over d\cos\theta} &=&\frac{\alpha _- \alpha _+ P_T^2 \left[(3\alpha_\psi +2) \cos ^2\theta +1\right]}{12 (\alpha_\psi +3)}  ,\nonumber\\
{d\langle \mu_2 \rangle\over d\cos\theta} &=&-\frac{\alpha _- \alpha _+ P_T^2 \gamma_\psi \sin ^2\theta  \cos ^2\theta}{6 (\alpha_\psi +3)}  ,\nonumber\\
{d\langle \mu_3 \rangle\over d\cos\theta} &=&-\frac{3 \alpha _+ P_T^2 \beta_\psi \sin ^2\theta }{8 (\alpha_\psi +3)}  ,\\
{d\langle \mu_4 \rangle\over d\cos\theta} &=&-\frac{3 \alpha _- P_T^2 \beta_\psi  \sin ^2\theta }{8 (\alpha_\psi +3)}  ,\nonumber\\
{d\langle \mu_5 \rangle\over d\cos\theta} &=&\frac{\alpha _- \alpha _+ P_T^2 \left[(2 \alpha_\psi +1) \cos ^2\theta+\alpha_\psi \right]}{12 (\alpha_\psi +3)} .\nonumber
\end{eqnarray}

\begin{figure}[ht]
    \centering
    \includegraphics[width=0.5\textwidth]{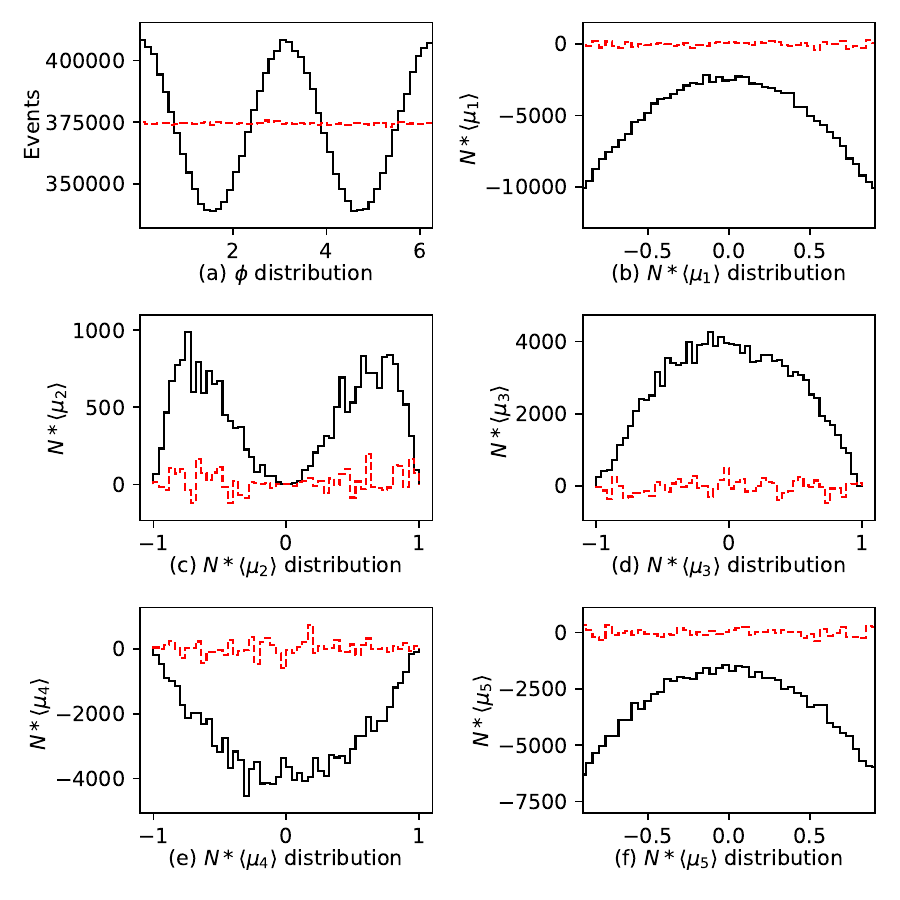}
    \caption{Various distributions in the $e^+e^-\to\Lambda\bar\Lambda$ process. (a). $\Lambda$ azimuthal angle distribution, (b-f): moments distributions of $N*\langle \mu_i\rangle (i=1,2,..,5)$, here $N$ is the number of toy Monte-Carlo events, using $e^+ e^- \to \psi(3686) \to \Lambda \overline\Lambda$ with $\alpha = 0.69$ and $\Delta\Phi = 23^\circ$ radian \cite{BESIII:2023euh}, and dashed histogram is for $P_T = 0$, histogram for $P_T=0.5$.}
    \label{fig:moments}
\end{figure}

The above moment analysis can be used to intuitively display the polarization information in the $e^+e^-\to\Lambda\bar\Lambda$ process and its transfer in the $\Lambda$ and $\bar\Lambda$ decay. In experiments, the observed variables corresponding to these moments are constructed using kinematic variables detected in the experiment, and the values are taken as the weight factors of each event in the distribution plot. If $P_T=0$, these moment distributions are trivially flat; if $P_T\neq0$, they should exhibit a nontrivial distribution described by the Eqs. (\ref{moments}). In order to compare with unweighted events in the experiment, we generate toy Monte-Carlo events using the joint angular distribution formula $\mathcal{W}(\xi)$ for the $e^+e^-\to\Lambda\bar\Lambda$ process, with parameters set to $\Delta\Phi=0.4$ rad, $\alpha_\psi=0.69$, ~$\alpha_-=0.748$ and $\alpha_+=-0.757$ \cite{BESIII:2023euh,BESIII:2018cnd}, and $P_T=0.5$. The $\phi$ in Fig. \ref{fig:moments} represents the azimuthal angle distribution of $\Lambda$, showing the distribution of $d\mathcal{W}(\xi)/d\phi\sim \alpha_\psi P_T^2 \cos(2\phi)$. The $\langle \mu_i\rangle(i=1,2,…,5)$ represents the nontrivial moment distributions, providing an intuitive display of the existence of beam transverse polarization. As a reference, the $P_t=0$ cases are also presented for comparison. It can be observed that they appear as flat distributions with some statistical fluctuations.

\subsection{Statistical significance test} \label{sec:significance}

Using the $J/\psi\to\Lambda\bar\Lambda$ decay, the BESIII collaboration has previously studied the decay parameters of $\Lambda$ and $\bar\Lambda$, and measured the precise values of $\alpha_-$ and $\alpha_+$ by fitting the data using the joint angular distribution formula for the 4-body decay. The formula used did not include the contribution of beam transverse polarization. At the $J/\psi$ energy point, due to the depolarization resonance effect, the transverse polarization of the beam becomes very small. However, at the $\psi(3686)$ energy point, the polarization effect of the beam will be more significant. Taking the $\psi(3686)\to\Lambda\bar\Lambda$ decay as an example, we elucidate the role of beam transverse polarization effects in measuring the decay parameters of $\Lambda$.

In experiments, the maximum likelihood method is commonly used to measure the decay parameters of $\Lambda$, and its statistical error can also be obtained from the fit to the experimental data. For the $\psi(3686)\to\Lambda\bar\Lambda, \Lambda\to p\pi^-, \bar\Lambda\to \bar p\pi^+$ decay, we define the probability distribution function that describes its joint angular distribution as:
\begin{equation}
\widetilde{\mathcal{W}} = {\mathcal{W}(\theta,\phi, \theta_1,\phi_1, \theta_2,\phi_2) \over \int \mathcal{W}(…)d\cos\theta d\cos\theta_1 d\cos\theta_2 d\phi d\phi_1 d\phi_2 }.
\end{equation}
Here, the denominator serves to normalize the probability distribution of the angular distribution. The likelihood function for the observed data sample of $N$ events in the experiment is expressed as:
\begin{equation}
\label{eq:lklhd}
L (\theta,\phi, \theta_1,\phi_1, \theta_2,\phi_2|\vec{\xi})= \prod_{i=1}^N \widetilde{\mathcal{W}}_i,
\end{equation}
here $\vec \xi=(\alpha_\psi, \Delta\Phi, \alpha_-, \alpha_+,P_T)$ are parameters to be estimated, and the product is computed based on the probability of the $i$-th event $\widetilde{\mathcal{W}}_i$. According to the maximum likelihood estimation method for parameter estimation, the precision of parameter $x_i$ is expressed as
\begin{equation}
\delta (x_i) = {\sqrt{V(x_i)}\over |x_i|},
\end{equation}
where $V(x_i)$ represents the variance of the parameter. We assume that the beam polarization $P_T$ corresponding to the $\psi(3686)$ data sample can be determined through other processes, such as $e^+e^-\to \mu^+\mu^-$ measurements. The maximum likelihood fit selects four parameters $\alpha_\psi, \Delta\Phi, \alpha_-$ and $\alpha_+$, and the error matrix formed by them can be calculated using the following equation:
\begin{equation}
V_{ij}^{-1}(\vec x)=N \int {1\over \widetilde{\mathcal{W}}} {\partial \widetilde{\mathcal{W}}\over \partial x_i }{\partial \widetilde{\mathcal{W}}\over \partial x_j } d\cos\theta d\cos\theta_1 d\cos\theta_2 d\phi d\phi d\phi_2.
\end{equation}

\begin{figure}[ht]
    \centering
    \includegraphics[width=0.5\textwidth]{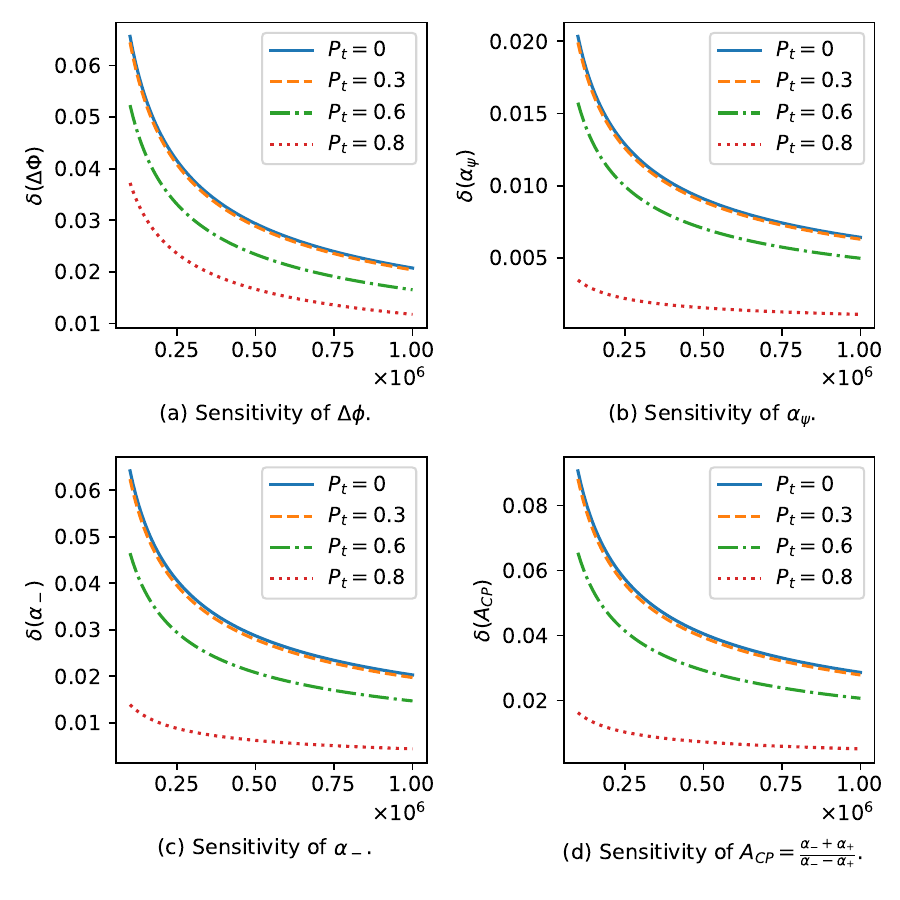}
    \caption{Sensitivity of $\Delta\Phi,~\alpha_\psi,~\alpha_-$ and $A_{CP} = (\alpha_- + \alpha_+)/(\alpha_- - \alpha_+)$ in terms of data size $N$,
    using $e^+ e^- \to \psi(3686) \to \Lambda \overline\Lambda$ with $\alpha_\psi = 0.69$, $\Delta\Phi = 23^\circ$ \cite{BESIII:2023euh}, and $\alpha_-=0.748$ and $\alpha_+=-0.757$ \cite{BESIII:2023euh,BESIII:2018cnd}. }
    \label{fig:sens}
\end{figure}

Figure \ref{fig:sens} shows the measurement sensitivity of estimated $\Delta\Phi,~\alpha_\psi,~\alpha_-$, and $A_{CP}={\alpha_-+\alpha_+\over \alpha_--\alpha_+}$ under different statistical events of $e^+e^- \to \psi(3686) \to \Lambda (p \pi^-) \bar\Lambda (\overline p \pi^+)$. 
In the absence of transverse beam polarization ($P_T=0$), the relative error in parameter measurement is maximized for the same data statistical quantity $N$. As the value of $P_T$ increases from 0.3 to 0.8, it can be observed that the measurement sensitivity of these parameters increases. In other words, the application of transverse beam polarization is advantageous for enhancing the measurement sensitivity of the parameters.

\begin{figure}[ht]
    \centering
    \includegraphics[width=0.5\textwidth]{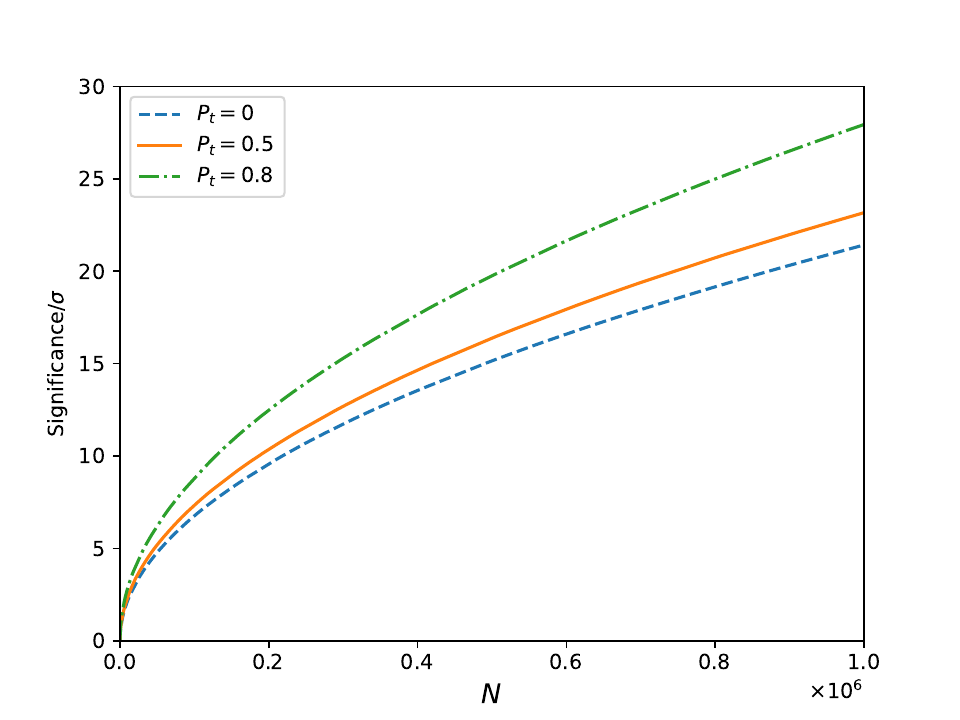}
    \caption{Significance test for \CP asymmetry in $\Lambda(\bar\Lambda)\to~p\pi^-(p\bar \pi^+)$ decays as a function of the number of observed events $N$, using toy Monte Carlo events for $e^+ e^- \to \psi(3686) \to \Lambda \overline\Lambda$ generated with parameters $\alpha_\psi = 0.69$, $\Delta\Phi = 23^\circ$ \cite{BESIII:2023euh}, and $\alpha_-=0.748$ and $\alpha_+=-0.757$~\cite{BESIII:2023euh,BESIII:2018cnd} for different transverse polarizations $P_t=0,~0.5$ and $0.8$.}
    \label{fig:sig}
\end{figure}

If $\Lambda(\bar\Lambda)\to~p\pi^-(p\bar \pi^+)$ decays conserve \CP symmetry, their decay parameters satisfy the relation $\alpha_-=-\alpha_+$. If an experiment measures $\alpha_-\neq-\alpha_+$, it implies \CP violation in $\Lambda(\bar\Lambda)$ decays. The significance test for \CP asymmetry can be attributed to statistical hypothesis testing as follows. The null hypothesis is that the sum of the $\Lambda$ and $\bar \Lambda$ decay parameters is zero, while the alternative hypothesis is that the sum is not zero. We conduct a test using toy Monte Carlo events generated with parameters $\alpha_\psi = 0.69$, $\Delta\Phi = 23^\circ$ \cite{BESIII:2023euh}, and $\alpha_-=0.748$ and $\alpha_+=-0.757$ \cite{BESIII:2023euh,BESIII:2018cnd} for different transverse polarizations $P_T=0,~0.5$ and $0.8$. The significance is calculated as $\sqrt{-2\ln~L_0-(-2\ln~L_1)}$, where $L_0$ and $L_1$ are the log-likelihood values for the null and alternative hypotheses, respectively. Using the likelihood function defined in Eq.(\ref{eq:lklhd}), $L_0$ is calculated with $\alpha_-=-\alpha_+=0.7525$, while $L_1$ is calculated with $\alpha_-=0.748$ and $\alpha_+=-0.757$ under different $P_T$ assumptions. The significance is shown in Fig. \ref{fig:sig} as a function of the number of observed events. It can be observed that the significance benefits from the non-zero transverse polarization of the $e^+e^-$ beams.

\section{Summary and Perspectives} \label{sec:close}

In summary, high-energy lepton beams naturally acquire transverse polarization in a storage ring through the mechanism of self-polarization, known as the Sokolov-Ternov effect.
We investigate the impact of transversely polarized beams on the observables of hyperon production and sequential decay at an electron-positron collider, utilizing helicity amplitude analysis.
It is shown that the transverse polarization of the beams introduces a new azimuthal modulation of the events in terms of the azimuthal angle of hyperons in the laboratory frame.
Through moments analysis and maximum likelihood estimation, we characterize the statistical uncertainties that transversely polarized beams may impose on constraining \CP violation parameters. As demonstrated in this paper, the sensitivity to measure the hyperon decay parameter and \CP violation can be enhanced by employing transversely polarized electron and positron beams.
In our study, we consider $e^+e^- \to \gamma^*/\psi \to \Lambda (p \pi^-) \bar\Lambda (\overline p \pi^+)$ and $\Xi^-(\Lambda \pi^-)\bar\Xi^+(\overline{\Lambda} \pi^+)$ as examples. 
However, the sensitivity remains largely unchanged if the transverse polarization of beam is small, e.g. in the case of $J/\psi$ decay at BESIII. Therefore, previous measurements of $J/\psi$ decays are not affected even if the transverse polarization of beams is considered. On the other hand, at the energies of big polarization degree, e.g. $\psi(3686)$, we recommend the inclusion of this effect in future data analyses at $e^+e^-$ circular colliders, and suggest extending the formalism in this paper to other processes such as $e^+e^-\to\Sigma^+ (p \pi^0) \bar{\Sigma}^- (\overline p \pi^0)$, $\Lambda_c^+ (\Lambda \pi^+) \overline\Lambda_c (\overline\Lambda \pi^-)$ and $\Omega^- (\Lambda K^-) \overline\Omega^+ (\overline\Lambda K^+)$ \cite{Zhang:2023wmd,Zhang:2023box} for comprehensive investigations.
This effect on other observables, such as higher-order quantum electrodynamic processes, hyperon weak radiative decays (e.g., \cite{Shi:2022dhw,Xing:2023jnr}), and hadronic vacuum polarization, is a topic of future interest as well.

\bigskip

\begin{acknowledgments}

We are grateful to Jianping Dai, Zhe Duan, Boxing Gou, Andrzej Kupsc, Xiongfei Wang, Jujun Xie, and Xiaorong Zhou for useful discussions. This work is supported by the National Key R\&D Program of China under Grant No. 2023YFA1606703, the National Natural Science Foundation of China  (Grants Nos.12075289, 12175244, U2032109) and the Strategic Priority Research Program of Chinese Academy of Sciences (Grant No. XDB34030301).

\end{acknowledgments}

\bibliography{NEFF}

\begin{thebibliography}{51}
\expandafter\ifx\csname natexlab\endcsname\relax\def\natexlab#1{#1}\fi
\expandafter\ifx\csname bibnamefont\endcsname\relax
  \def\bibnamefont#1{#1}\fi
\expandafter\ifx\csname bibfnamefont\endcsname\relax
  \def\bibfnamefont#1{#1}\fi
\expandafter\ifx\csname citenamefont\endcsname\relax
  \def\citenamefont#1{#1}\fi
\expandafter\ifx\csname url\endcsname\relax
  \def\url#1{\texttt{#1}}\fi
\expandafter\ifx\csname urlprefix\endcsname\relax\def\urlprefix{URL }\fi
\providecommand{\bibinfo}[2]{#2}
\providecommand{\eprint}[2][]{\url{#2}}

\bibitem[{\citenamefont{Ananthanarayan and
  Rindani}(2004)}]{Ananthanarayan:2003wi}
\bibinfo{author}{\bibfnamefont{B.}~\bibnamefont{Ananthanarayan}}
  \bibnamefont{and} \bibinfo{author}{\bibfnamefont{S.~D.}
  \bibnamefont{Rindani}}, \bibinfo{journal}{Phys. Rev. D}
  \textbf{\bibinfo{volume}{70}}, \bibinfo{pages}{036005}
  (\bibinfo{year}{2004}), \eprint{hep-ph/0309260}.

\bibitem[{\citenamefont{Moortgat-Pick et~al.}(2008)}]{Moortgat-Pick:2005jsx}
\bibinfo{author}{\bibfnamefont{G.}~\bibnamefont{Moortgat-Pick}}
  \bibnamefont{et~al.}, \bibinfo{journal}{Phys. Rept.}
  \textbf{\bibinfo{volume}{460}}, \bibinfo{pages}{131} (\bibinfo{year}{2008}),
  \eprint{hep-ph/0507011}.

\bibitem[{\citenamefont{Wen et~al.}(2023)\citenamefont{Wen, Yan, Yu, and
  Yuan}}]{Wen:2023xxc}
\bibinfo{author}{\bibfnamefont{X.-K.} \bibnamefont{Wen}},
  \bibinfo{author}{\bibfnamefont{B.}~\bibnamefont{Yan}},
  \bibinfo{author}{\bibfnamefont{Z.}~\bibnamefont{Yu}}, \bibnamefont{and}
  \bibinfo{author}{\bibfnamefont{C.~P.} \bibnamefont{Yuan}},
  \bibinfo{journal}{Phys. Rev. Lett.} \textbf{\bibinfo{volume}{131}},
  \bibinfo{pages}{241801} (\bibinfo{year}{2023}), \eprint{2307.05236}.

\bibitem[{\citenamefont{Chen et~al.}(2022)}]{Chen:2022rgo}
\bibinfo{author}{\bibfnamefont{S.~H.} \bibnamefont{Chen}} \bibnamefont{et~al.},
  \bibinfo{journal}{JINST} \textbf{\bibinfo{volume}{17}},
  \bibinfo{pages}{P08005} (\bibinfo{year}{2022}).

\bibitem[{\citenamefont{Sokolov and Ternov}(1963)}]{Sokolov:1963zn}
\bibinfo{author}{\bibfnamefont{A.~A.} \bibnamefont{Sokolov}} \bibnamefont{and}
  \bibinfo{author}{\bibfnamefont{I.~M.} \bibnamefont{Ternov}},
  \bibinfo{journal}{Dokl. Akad. Nauk SSSR} \textbf{\bibinfo{volume}{153}},
  \bibinfo{pages}{1052} (\bibinfo{year}{1963}).

\bibitem[{\citenamefont{Baier and Fadin}(1965)}]{Baier:1965po}
\bibinfo{author}{\bibfnamefont{V.~N.} \bibnamefont{Baier}} \bibnamefont{and}
  \bibinfo{author}{\bibfnamefont{V.~S.} \bibnamefont{Fadin}},
  \bibinfo{journal}{Sov. Phys. Dokl.} \textbf{\bibinfo{volume}{10}},
  \bibinfo{pages}{204} (\bibinfo{year}{1965}).

\bibitem[{\citenamefont{Jackson}(1976)}]{Jackson:1975qi}
\bibinfo{author}{\bibfnamefont{J.~D.} \bibnamefont{Jackson}},
  \bibinfo{journal}{Rev. Mod. Phys.} \textbf{\bibinfo{volume}{48}},
  \bibinfo{pages}{417} (\bibinfo{year}{1976}).

\bibitem[{\citenamefont{Montague}(1984)}]{Montague:1983yi}
\bibinfo{author}{\bibfnamefont{B.~W.} \bibnamefont{Montague}},
  \bibinfo{journal}{Phys. Rept.} \textbf{\bibinfo{volume}{113}},
  \bibinfo{pages}{1} (\bibinfo{year}{1984}).

\bibitem[{\citenamefont{Kurdadze et~al.}(1976)\citenamefont{Kurdadze,
  Serednyakov, Sidorov, Skrinsky, Tumaikin, and Shatunov}}]{Kurdadze:1975zm}
\bibinfo{author}{\bibfnamefont{L.~M.} \bibnamefont{Kurdadze}},
  \bibinfo{author}{\bibfnamefont{S.~I.} \bibnamefont{Serednyakov}},
  \bibinfo{author}{\bibfnamefont{V.~A.} \bibnamefont{Sidorov}},
  \bibinfo{author}{\bibfnamefont{A.~N.} \bibnamefont{Skrinsky}},
  \bibinfo{author}{\bibfnamefont{G.~M.} \bibnamefont{Tumaikin}},
  \bibnamefont{and} \bibinfo{author}{\bibfnamefont{Y.~M.}
  \bibnamefont{Shatunov}}, \bibinfo{journal}{Sov. Phys. JETP}
  \textbf{\bibinfo{volume}{44}}, \bibinfo{pages}{1063} (\bibinfo{year}{1976}).

\bibitem[{\citenamefont{Learned et~al.}(1975)\citenamefont{Learned, Resvanis,
  and Spencer}}]{Learned:1975sw}
\bibinfo{author}{\bibfnamefont{J.~G.} \bibnamefont{Learned}},
  \bibinfo{author}{\bibfnamefont{L.~K.} \bibnamefont{Resvanis}},
  \bibnamefont{and} \bibinfo{author}{\bibfnamefont{C.~M.}
  \bibnamefont{Spencer}}, \bibinfo{journal}{Phys. Rev. Lett.}
  \textbf{\bibinfo{volume}{35}}, \bibinfo{pages}{1688} (\bibinfo{year}{1975}).

\bibitem[{\citenamefont{Knudsen et~al.}(1991)\citenamefont{Knudsen, Koutchouk,
  Placidi, Schmidt, Crozon, Badier, Blondel, and Dehning}}]{Knudsen:1991cu}
\bibinfo{author}{\bibfnamefont{L.}~\bibnamefont{Knudsen}},
  \bibinfo{author}{\bibfnamefont{J.~P.} \bibnamefont{Koutchouk}},
  \bibinfo{author}{\bibfnamefont{M.}~\bibnamefont{Placidi}},
  \bibinfo{author}{\bibfnamefont{R.}~\bibnamefont{Schmidt}},
  \bibinfo{author}{\bibfnamefont{M.}~\bibnamefont{Crozon}},
  \bibinfo{author}{\bibfnamefont{J.}~\bibnamefont{Badier}},
  \bibinfo{author}{\bibfnamefont{A.}~\bibnamefont{Blondel}}, \bibnamefont{and}
  \bibinfo{author}{\bibfnamefont{B.}~\bibnamefont{Dehning}},
  \bibinfo{journal}{Phys. Lett. B} \textbf{\bibinfo{volume}{270}},
  \bibinfo{pages}{97} (\bibinfo{year}{1991}).

\bibitem[{\citenamefont{Nikitin}(2019)}]{Nikitin:2019nxl}
\bibinfo{author}{\bibfnamefont{S.}~\bibnamefont{Nikitin}},
  \bibinfo{journal}{Int. J. Mod. Phys. A} \textbf{\bibinfo{volume}{34}},
  \bibinfo{pages}{1940004} (\bibinfo{year}{2019}).

\bibitem[{\citenamefont{Blondel et~al.}(2019)}]{Blondel:2019jmp}
\bibinfo{author}{\bibfnamefont{A.}~\bibnamefont{Blondel}} \bibnamefont{et~al.}
  (\bibinfo{year}{2019}), \eprint{1909.12245}.

\bibitem[{\citenamefont{Asner et~al.}(2022)}]{USBelleIIGroup:2022qro}
\bibinfo{author}{\bibfnamefont{D.~M.} \bibnamefont{Asner}} \bibnamefont{et~al.}
  (\bibinfo{collaboration}{US Belle II Group, Belle II/SuperKEKB e-
  Polarization Upgrade Working Group}), in \emph{\bibinfo{booktitle}{{Snowmass
  2021}}} (\bibinfo{year}{2022}), \eprint{2205.12847}.

\bibitem[{\citenamefont{Achasov et~al.}(2024)}]{Achasov:2023gey}
\bibinfo{author}{\bibfnamefont{M.}~\bibnamefont{Achasov}} \bibnamefont{et~al.},
  \bibinfo{journal}{Front. Phys. (Beijing)} \textbf{\bibinfo{volume}{19}},
  \bibinfo{pages}{14701} (\bibinfo{year}{2024}), \eprint{2303.15790}.

\bibitem[{\citenamefont{Czyz et~al.}(2007)\citenamefont{Czyz, Grzelinska, and
  Kuhn}}]{Czyz:2007wi}
\bibinfo{author}{\bibfnamefont{H.}~\bibnamefont{Czyz}},
  \bibinfo{author}{\bibfnamefont{A.}~\bibnamefont{Grzelinska}},
  \bibnamefont{and} \bibinfo{author}{\bibfnamefont{J.~H.} \bibnamefont{Kuhn}},
  \bibinfo{journal}{Phys. Rev. D} \textbf{\bibinfo{volume}{75}},
  \bibinfo{pages}{074026} (\bibinfo{year}{2007}), \eprint{hep-ph/0702122}.

\bibitem[{\citenamefont{Chen and Ping}(2007)}]{Chen:2007zzf}
\bibinfo{author}{\bibfnamefont{H.}~\bibnamefont{Chen}} \bibnamefont{and}
  \bibinfo{author}{\bibfnamefont{R.-G.} \bibnamefont{Ping}},
  \bibinfo{journal}{Phys. Rev. D} \textbf{\bibinfo{volume}{76}},
  \bibinfo{pages}{036005} (\bibinfo{year}{2007}).

\bibitem[{\citenamefont{F\"aldt and Kupsc}(2017)}]{Faldt:2017kgy}
\bibinfo{author}{\bibfnamefont{G.}~\bibnamefont{F\"aldt}} \bibnamefont{and}
  \bibinfo{author}{\bibfnamefont{A.}~\bibnamefont{Kupsc}},
  \bibinfo{journal}{Phys. Lett. B} \textbf{\bibinfo{volume}{772}},
  \bibinfo{pages}{16} (\bibinfo{year}{2017}), \eprint{1702.07288}.

\bibitem[{\citenamefont{Lee and Yang}(1957)}]{Lee:1957qs}
\bibinfo{author}{\bibfnamefont{T.~D.} \bibnamefont{Lee}} \bibnamefont{and}
  \bibinfo{author}{\bibfnamefont{C.-N.} \bibnamefont{Yang}},
  \bibinfo{journal}{Phys. Rev.} \textbf{\bibinfo{volume}{108}},
  \bibinfo{pages}{1645} (\bibinfo{year}{1957}).

\bibitem[{\citenamefont{Donoghue and Pakvasa}(1985)}]{Donoghue:1985ww}
\bibinfo{author}{\bibfnamefont{J.~F.} \bibnamefont{Donoghue}} \bibnamefont{and}
  \bibinfo{author}{\bibfnamefont{S.}~\bibnamefont{Pakvasa}},
  \bibinfo{journal}{Phys. Rev. Lett.} \textbf{\bibinfo{volume}{55}},
  \bibinfo{pages}{162} (\bibinfo{year}{1985}).

\bibitem[{\citenamefont{Donoghue et~al.}(1986)\citenamefont{Donoghue, He, and
  Pakvasa}}]{Donoghue:1986hh}
\bibinfo{author}{\bibfnamefont{J.~F.} \bibnamefont{Donoghue}},
  \bibinfo{author}{\bibfnamefont{X.-G.} \bibnamefont{He}}, \bibnamefont{and}
  \bibinfo{author}{\bibfnamefont{S.}~\bibnamefont{Pakvasa}},
  \bibinfo{journal}{Phys. Rev. D} \textbf{\bibinfo{volume}{34}},
  \bibinfo{pages}{833} (\bibinfo{year}{1986}).

\bibitem[{\citenamefont{Salone et~al.}(2022)\citenamefont{Salone, Adlarson,
  Batozskaya, Kupsc, Leupold, and Tandean}}]{Salone:2022lpt}
\bibinfo{author}{\bibfnamefont{N.}~\bibnamefont{Salone}},
  \bibinfo{author}{\bibfnamefont{P.}~\bibnamefont{Adlarson}},
  \bibinfo{author}{\bibfnamefont{V.}~\bibnamefont{Batozskaya}},
  \bibinfo{author}{\bibfnamefont{A.}~\bibnamefont{Kupsc}},
  \bibinfo{author}{\bibfnamefont{S.}~\bibnamefont{Leupold}}, \bibnamefont{and}
  \bibinfo{author}{\bibfnamefont{J.}~\bibnamefont{Tandean}},
  \bibinfo{journal}{Phys. Rev. D} \textbf{\bibinfo{volume}{105}},
  \bibinfo{pages}{116022} (\bibinfo{year}{2022}), \eprint{2203.03035}.

\bibitem[{\citenamefont{Zeng et~al.}(2023)\citenamefont{Zeng, Xu, Zhou, Qin,
  and Zheng}}]{Zeng:2023wqw}
\bibinfo{author}{\bibfnamefont{S.}~\bibnamefont{Zeng}},
  \bibinfo{author}{\bibfnamefont{Y.}~\bibnamefont{Xu}},
  \bibinfo{author}{\bibfnamefont{X.~R.} \bibnamefont{Zhou}},
  \bibinfo{author}{\bibfnamefont{J.~J.} \bibnamefont{Qin}}, \bibnamefont{and}
  \bibinfo{author}{\bibfnamefont{B.}~\bibnamefont{Zheng}},
  \bibinfo{journal}{Chin. Phys. C} \textbf{\bibinfo{volume}{47}},
  \bibinfo{pages}{113001} (\bibinfo{year}{2023}), \eprint{2306.15602}.

\bibitem[{\citenamefont{Dubnickova et~al.}(1996)\citenamefont{Dubnickova,
  Dubnicka, and Rekalo}}]{Dubnickova:1992ii}
\bibinfo{author}{\bibfnamefont{A.~Z.} \bibnamefont{Dubnickova}},
  \bibinfo{author}{\bibfnamefont{S.}~\bibnamefont{Dubnicka}}, \bibnamefont{and}
  \bibinfo{author}{\bibfnamefont{M.~P.} \bibnamefont{Rekalo}},
  \bibinfo{journal}{Nuovo Cim. A} \textbf{\bibinfo{volume}{109}},
  \bibinfo{pages}{241} (\bibinfo{year}{1996}).

\bibitem[{\citenamefont{Brodsky et~al.}(2004)\citenamefont{Brodsky, Carlson,
  Hiller, and Hwang}}]{Brodsky:2003gs}
\bibinfo{author}{\bibfnamefont{S.~J.} \bibnamefont{Brodsky}},
  \bibinfo{author}{\bibfnamefont{C.~E.} \bibnamefont{Carlson}},
  \bibinfo{author}{\bibfnamefont{J.~R.} \bibnamefont{Hiller}},
  \bibnamefont{and} \bibinfo{author}{\bibfnamefont{D.~S.} \bibnamefont{Hwang}},
  \bibinfo{journal}{Phys. Rev. D} \textbf{\bibinfo{volume}{69}},
  \bibinfo{pages}{054022} (\bibinfo{year}{2004}), \eprint{hep-ph/0310277}.

\bibitem[{\citenamefont{Tomasi-Gustafsson
  et~al.}(2005)\citenamefont{Tomasi-Gustafsson, Lacroix, Duterte, and
  Gakh}}]{Tomasi-Gustafsson:2005svz}
\bibinfo{author}{\bibfnamefont{E.}~\bibnamefont{Tomasi-Gustafsson}},
  \bibinfo{author}{\bibfnamefont{F.}~\bibnamefont{Lacroix}},
  \bibinfo{author}{\bibfnamefont{C.}~\bibnamefont{Duterte}}, \bibnamefont{and}
  \bibinfo{author}{\bibfnamefont{G.~I.} \bibnamefont{Gakh}},
  \bibinfo{journal}{Eur. Phys. J. A} \textbf{\bibinfo{volume}{24}},
  \bibinfo{pages}{419} (\bibinfo{year}{2005}), \eprint{nucl-th/0503001}.

\bibitem[{\citenamefont{Tsai}(1975)}]{Tsai:1975bd}
\bibinfo{author}{\bibfnamefont{Y.-S.} \bibnamefont{Tsai}},
  \bibinfo{journal}{Phys. Rev. D} \textbf{\bibinfo{volume}{12}},
  \bibinfo{pages}{3533} (\bibinfo{year}{1975}).

\bibitem[{\citenamefont{Bletzacker and Nieh}(1976)}]{Bletzacker:1976npa}
\bibinfo{author}{\bibfnamefont{F.}~\bibnamefont{Bletzacker}} \bibnamefont{and}
  \bibinfo{author}{\bibfnamefont{H.~T.} \bibnamefont{Nieh}},
  \bibinfo{journal}{Phys. Rev. D} \textbf{\bibinfo{volume}{14}},
  \bibinfo{pages}{1251} (\bibinfo{year}{1976}).

\bibitem[{\citenamefont{Hikasa}(1986)}]{Hikasa:1985qi}
\bibinfo{author}{\bibfnamefont{K.-i.} \bibnamefont{Hikasa}},
  \bibinfo{journal}{Phys. Rev. D} \textbf{\bibinfo{volume}{33}},
  \bibinfo{pages}{3203} (\bibinfo{year}{1986}).

\bibitem[{\citenamefont{Ablikim et~al.}(2023{\natexlab{a}})}]{BESIII:2023euh}
\bibinfo{author}{\bibfnamefont{M.}~\bibnamefont{Ablikim}} \bibnamefont{et~al.}
  (\bibinfo{collaboration}{BESIII}), \bibinfo{journal}{JHEP}
  \textbf{\bibinfo{volume}{10}}, \bibinfo{pages}{081}
  (\bibinfo{year}{2023}{\natexlab{a}}), \bibinfo{note}{[Erratum: JHEP 12, 080
  (2023)]}, \eprint{2303.00271}.

\bibitem[{\citenamefont{Ablikim et~al.}(2020)}]{BESIII:2020fqg}
\bibinfo{author}{\bibfnamefont{M.}~\bibnamefont{Ablikim}} \bibnamefont{et~al.}
  (\bibinfo{collaboration}{BESIII}), \bibinfo{journal}{Phys. Rev. Lett.}
  \textbf{\bibinfo{volume}{125}}, \bibinfo{pages}{052004}
  (\bibinfo{year}{2020}), \eprint{2004.07701}.

\bibitem[{\citenamefont{Ablikim et~al.}(2022{\natexlab{a}})}]{BESIII:2021cvv}
\bibinfo{author}{\bibfnamefont{M.}~\bibnamefont{Ablikim}} \bibnamefont{et~al.}
  (\bibinfo{collaboration}{BESIII}), \bibinfo{journal}{Phys. Rev. D}
  \textbf{\bibinfo{volume}{105}}, \bibinfo{pages}{L011101}
  (\bibinfo{year}{2022}{\natexlab{a}}), \eprint{2111.11742}.

\bibitem[{\citenamefont{Ablikim et~al.}(2022{\natexlab{b}})}]{BESIII:2022lsz}
\bibinfo{author}{\bibfnamefont{M.}~\bibnamefont{Ablikim}} \bibnamefont{et~al.}
  (\bibinfo{collaboration}{BESIII}), \bibinfo{journal}{Phys. Rev. D}
  \textbf{\bibinfo{volume}{106}}, \bibinfo{pages}{L091101}
  (\bibinfo{year}{2022}{\natexlab{b}}), \eprint{2206.10900}.

\bibitem[{\citenamefont{Liu et~al.}(2023)\citenamefont{Liu, Zhang, and
  Wang}}]{Liu:2023xhg}
\bibinfo{author}{\bibfnamefont{H.}~\bibnamefont{Liu}},
  \bibinfo{author}{\bibfnamefont{J.}~\bibnamefont{Zhang}}, \bibnamefont{and}
  \bibinfo{author}{\bibfnamefont{X.}~\bibnamefont{Wang}},
  \bibinfo{journal}{Symmetry} \textbf{\bibinfo{volume}{15}},
  \bibinfo{pages}{214} (\bibinfo{year}{2023}).

\bibitem[{\citenamefont{Ablikim et~al.}(2023{\natexlab{b}})}]{BESIII:2023lkg}
\bibinfo{author}{\bibfnamefont{M.}~\bibnamefont{Ablikim}} \bibnamefont{et~al.}
  (\bibinfo{collaboration}{BESIII}), \bibinfo{journal}{Phys. Rev. D}
  \textbf{\bibinfo{volume}{108}}, \bibinfo{pages}{L011101}
  (\bibinfo{year}{2023}{\natexlab{b}}), \eprint{2302.09767}.

\bibitem[{\citenamefont{Ablikim et~al.}(2019{\natexlab{a}})}]{BESIII:2019dve}
\bibinfo{author}{\bibfnamefont{M.}~\bibnamefont{Ablikim}} \bibnamefont{et~al.}
  (\bibinfo{collaboration}{BESIII}), \bibinfo{journal}{Phys. Rev. D}
  \textbf{\bibinfo{volume}{100}}, \bibinfo{pages}{051101}
  (\bibinfo{year}{2019}{\natexlab{a}}), \eprint{1907.13041}.

\bibitem[{\citenamefont{Ablikim et~al.}(2021)}]{BESIII:2020lkm}
\bibinfo{author}{\bibfnamefont{M.}~\bibnamefont{Ablikim}} \bibnamefont{et~al.}
  (\bibinfo{collaboration}{BESIII}), \bibinfo{journal}{Phys. Rev. Lett.}
  \textbf{\bibinfo{volume}{126}}, \bibinfo{pages}{092002}
  (\bibinfo{year}{2021}), \eprint{2007.03679}.

\bibitem[{\citenamefont{Dai et~al.}(2023)\citenamefont{Dai, Cao, and
  Lenske}}]{Dai:2023vsw}
\bibinfo{author}{\bibfnamefont{J.-P.} \bibnamefont{Dai}},
  \bibinfo{author}{\bibfnamefont{X.}~\bibnamefont{Cao}}, \bibnamefont{and}
  \bibinfo{author}{\bibfnamefont{H.}~\bibnamefont{Lenske}},
  \bibinfo{journal}{Phys. Lett. B} \textbf{\bibinfo{volume}{846}},
  \bibinfo{pages}{138192} (\bibinfo{year}{2023}), \eprint{2304.04913}.

\bibitem[{\citenamefont{Perotti et~al.}(2019)\citenamefont{Perotti, F\"aldt,
  Kupsc, Leupold, and Song}}]{Perotti:2018wxm}
\bibinfo{author}{\bibfnamefont{E.}~\bibnamefont{Perotti}},
  \bibinfo{author}{\bibfnamefont{G.}~\bibnamefont{F\"aldt}},
  \bibinfo{author}{\bibfnamefont{A.}~\bibnamefont{Kupsc}},
  \bibinfo{author}{\bibfnamefont{S.}~\bibnamefont{Leupold}}, \bibnamefont{and}
  \bibinfo{author}{\bibfnamefont{J.~J.} \bibnamefont{Song}},
  \bibinfo{journal}{Phys. Rev. D} \textbf{\bibinfo{volume}{99}},
  \bibinfo{pages}{056008} (\bibinfo{year}{2019}), \eprint{1809.04038}.

\bibitem[{\citenamefont{Bai et~al.}(1995)}]{BES:1995wyo}
\bibinfo{author}{\bibfnamefont{J.~Z.} \bibnamefont{Bai}} \bibnamefont{et~al.}
  (\bibinfo{collaboration}{BES}), \bibinfo{journal}{Phys. Lett. B}
  \textbf{\bibinfo{volume}{355}}, \bibinfo{pages}{374} (\bibinfo{year}{1995}),
  \bibinfo{note}{[Erratum: Phys.Lett.B 363, 267 (1995)]}.

\bibitem[{\citenamefont{Ablikim et~al.}(2019{\natexlab{b}})}]{BESIII:2018wid}
\bibinfo{author}{\bibfnamefont{M.}~\bibnamefont{Ablikim}} \bibnamefont{et~al.}
  (\bibinfo{collaboration}{BESIII}), \bibinfo{journal}{Phys. Lett. B}
  \textbf{\bibinfo{volume}{791}}, \bibinfo{pages}{375}
  (\bibinfo{year}{2019}{\natexlab{b}}), \eprint{1808.02166}.

\bibitem[{\citenamefont{Anashin et~al.}(2018)}]{Anashin:2018iwp}
\bibinfo{author}{\bibfnamefont{V.~V.} \bibnamefont{Anashin}}
  \bibnamefont{et~al.}, \bibinfo{journal}{Phys. Lett. B}
  \textbf{\bibinfo{volume}{781}}, \bibinfo{pages}{174} (\bibinfo{year}{2018}),
  \eprint{1801.10362}.

\bibitem[{\citenamefont{Hong et~al.}(2023)\citenamefont{Hong, Ping, Luo, Zhou,
  and Li}}]{Hong:2023soc}
\bibinfo{author}{\bibfnamefont{P.-C.} \bibnamefont{Hong}},
  \bibinfo{author}{\bibfnamefont{R.-G.} \bibnamefont{Ping}},
  \bibinfo{author}{\bibfnamefont{T.}~\bibnamefont{Luo}},
  \bibinfo{author}{\bibfnamefont{X.-R.} \bibnamefont{Zhou}}, \bibnamefont{and}
  \bibinfo{author}{\bibfnamefont{H.}~\bibnamefont{Li}}, \bibinfo{journal}{Chin.
  Phys. C} \textbf{\bibinfo{volume}{47}}, \bibinfo{pages}{093103}
  (\bibinfo{year}{2023}), \eprint{2306.08517}.

\bibitem[{\citenamefont{Chen and Ping}(2019)}]{Chen:2019hqi}
\bibinfo{author}{\bibfnamefont{H.}~\bibnamefont{Chen}} \bibnamefont{and}
  \bibinfo{author}{\bibfnamefont{R.-G.} \bibnamefont{Ping}},
  \bibinfo{journal}{Phys. Rev. D} \textbf{\bibinfo{volume}{99}},
  \bibinfo{pages}{114027} (\bibinfo{year}{2019}).

\bibitem[{\citenamefont{F\"aldt}(2018)}]{Faldt:2017yqt}
\bibinfo{author}{\bibfnamefont{G.}~\bibnamefont{F\"aldt}},
  \bibinfo{journal}{Phys. Rev. D} \textbf{\bibinfo{volume}{97}},
  \bibinfo{pages}{053002} (\bibinfo{year}{2018}), \eprint{1709.01803}.

\bibitem[{\citenamefont{Ablikim et~al.}(2019{\natexlab{c}})}]{BESIII:2019odb}
\bibinfo{author}{\bibfnamefont{M.}~\bibnamefont{Ablikim}} \bibnamefont{et~al.}
  (\bibinfo{collaboration}{BESIII}), \bibinfo{journal}{Phys. Rev. D}
  \textbf{\bibinfo{volume}{100}}, \bibinfo{pages}{072004}
  (\bibinfo{year}{2019}{\natexlab{c}}), \eprint{1905.04707}.

\bibitem[{\citenamefont{Ablikim et~al.}(2019{\natexlab{d}})}]{BESIII:2018cnd}
\bibinfo{author}{\bibfnamefont{M.}~\bibnamefont{Ablikim}} \bibnamefont{et~al.}
  (\bibinfo{collaboration}{BESIII}), \bibinfo{journal}{Nature Phys.}
  \textbf{\bibinfo{volume}{15}}, \bibinfo{pages}{631}
  (\bibinfo{year}{2019}{\natexlab{d}}), \eprint{1808.08917}.

\bibitem[{\citenamefont{Zhang and Song}(2023)}]{Zhang:2023wmd}
\bibinfo{author}{\bibfnamefont{Z.}~\bibnamefont{Zhang}} \bibnamefont{and}
  \bibinfo{author}{\bibfnamefont{J.~J.} \bibnamefont{Song}},
  \bibinfo{journal}{Chin. Phys. C} \textbf{\bibinfo{volume}{47}},
  \bibinfo{pages}{093101} (\bibinfo{year}{2023}), \eprint{2303.02629}.

\bibitem[{\citenamefont{Zhang et~al.}(2024)\citenamefont{Zhang, Song, and
  Zhou}}]{Zhang:2023box}
\bibinfo{author}{\bibfnamefont{Z.}~\bibnamefont{Zhang}},
  \bibinfo{author}{\bibfnamefont{J.~J.} \bibnamefont{Song}}, \bibnamefont{and}
  \bibinfo{author}{\bibfnamefont{Y.-j.} \bibnamefont{Zhou}},
  \bibinfo{journal}{Phys. Rev. D} \textbf{\bibinfo{volume}{109}},
  \bibinfo{pages}{036005} (\bibinfo{year}{2024}), \eprint{2312.04363}.

\bibitem[{\citenamefont{Shi et~al.}(2022)\citenamefont{Shi, Li, Lu, and
  Geng}}]{Shi:2022dhw}
\bibinfo{author}{\bibfnamefont{R.-X.} \bibnamefont{Shi}},
  \bibinfo{author}{\bibfnamefont{S.-Y.} \bibnamefont{Li}},
  \bibinfo{author}{\bibfnamefont{J.-X.} \bibnamefont{Lu}}, \bibnamefont{and}
  \bibinfo{author}{\bibfnamefont{L.-S.} \bibnamefont{Geng}},
  \bibinfo{journal}{Sci. Bull.} \textbf{\bibinfo{volume}{67}},
  \bibinfo{pages}{2298} (\bibinfo{year}{2022}), \eprint{2206.11773}.

\bibitem[{\citenamefont{Xing et~al.}(2023)\citenamefont{Xing, Shi, Sun, and
  Zhao}}]{Xing:2023jnr}
\bibinfo{author}{\bibfnamefont{Z.-P.} \bibnamefont{Xing}},
  \bibinfo{author}{\bibfnamefont{Y.~J.} \bibnamefont{Shi}},
  \bibinfo{author}{\bibfnamefont{J.}~\bibnamefont{Sun}}, \bibnamefont{and}
  \bibinfo{author}{\bibfnamefont{Z.-X.} \bibnamefont{Zhao}}
  (\bibinfo{year}{2023}), \eprint{2312.17568}.

\end{thebibliography}

\end{document}